\documentclass[%
 preprint,
 amsmath,amssymb,
 aps,
pre,
]{revtex4-2}

\usepackage{graphicx}% Include figure files
\usepackage{dcolumn}% Align table columns on decimal point
\usepackage{bm}% bold math
\usepackage{hyperref}% add hypertext capabilities
\usepackage[mathlines]{lineno}% Enable numbering of text and display math
\begin{document}

\title{Density-Selected Topological Pathways in the Melting of
Single-Particle-Thick Stripes}% Force line breaks with \\
\author{José Rafael Bordin}
\affiliation{Fachbereich Physik, Universität Konstanz, Konstanz, Deutschland}
 \affiliation{Departamento de Física, Instituto de Física e Matemática, Universidade Federal de Pelotas, Pelotas, RS, Brasil}%Lines break automatically or can be forced with \\
\email{jrbordin@ufpel.edu.br}

\date{\today}% It is always \today, today,
             %  but any date may be explicitly specified

\begin{abstract}
The melting of stripe-forming systems involves changes in connectivity that
are not fully captured by conventional structural and orientational
descriptors. We investigate a two-dimensional model with competing
interactions whose low-temperature phase consists of one-particle-thick
stripes. Molecular dynamics simulations along seven heating isochores are
analyzed using thermodynamic, orientational, dynamical, and graph-based
observables. Heating produces a multistage reconstruction in which the loss
of stripe alignment and the reorganization of filament connectivity occur
over distinct temperature ranges. Density controls whether the disordered
filaments fragment into finite polymer-like clusters or remain joined in a
dynamically fluid, system-spanning network. The topological observables
distinguish these outcomes, which are not resolved by the thermodynamic and
orientational responses alone. Thus, the same ordered stripe microphase can
melt into topologically distinct fluids selected by density.

\end{abstract}
%\keywords{Suggested keywords}%Use showkeys class option if keyword
                              %display desired
\maketitle

%\tableofcontents

\section{Introduction}
\label{sec:introduction}

Competing interactions provide an economical route from isotropic
building blocks to spatially modulated matter. When aggregation or local
packing at one length scale is frustrated by a second interaction scale,
macroscopic phase separation can be replaced by finite-wavelength
organization into clusters, stripes, lamellae, bubbles, and interconnected
networks \cite{SeulAndelman1995,CiachPekalskiGozdz2013,
ZhuangZhangCharbonneau2016,Royall2018,LiuXi2019}. This mechanism appears in
systems as diverse as magnetic films, Langmuir monolayers, block copolymers,
protein solutions, and colloidal suspensions. In colloids, this phenomenology
is commonly associated with short-range attraction and long-range repulsion
(SALR), but modulated structures can also arise from purely repulsive
core-softened potentials possessing two competing particle separations.
Early studies by Jagla showed that core collapse in two dimensions can be
preempted by nontriangular ground states, including chains, squares,
hexagonal motifs, and quasicrystalline arrangements
\cite{Jagla1998CoreCollapse,Jagla1999MinimumEnergy}. Subsequent studies
established stripe and lamellar formation as generic consequences of
isotropic repulsive length-scale competition
\cite{MalescioPellicane2003,Glaser2007Mesophases,
PattabhiramanDijkstra2017,Somerville2020}. Such isotropic models are therefore
minimal platforms for separating the generic consequences of length-scale
competition from chemical or particle-specific details.

The phase behavior generated by competing interactions is by now well
established theoretically and computationally. Mesoscopic theories connect
SALR fluids with Brazovskii-type free-energy functionals and predict a generic
sequence of periodic morphologies
\cite{Ciach2008,CiachPekalskiGozdz2013}. Particle-resolved simulations have
identified equilibrium and metastable cluster fluids, lamellae, cylindrical
domains, percolated networks, and gyroid-like phases
\cite{ImperioReatto2006,ArcherWilding2007,
ZhuangZhangCharbonneau2016,EdelmannRoth2016,Bordin2023}. In two dimensions, where
thermal fluctuations and defects are especially important, stripes can form
either from competing attraction and repulsion or from an isotropic repulsive
shoulder
\cite{MalescioPellicane2003,ImperioReatto2006,
McDermottReichhardt2014}. Recent work has further shown that changes in
composition, confinement, substrate modulation, and particle anisotropy can
select among straight, modulated, labyrinthine, and fragmented stripe
patterns
\cite{PekalskiCiach2018,DeVirgiliis2024,
BordinPatterns2022,Reichhardt2024}. These results demonstrate that a stripe
is not specified solely by a peak in the structure factor: its thickness,
continuity, curvature, branching, and global connectivity constitute
independent structural degrees of freedom.

Experimental realizations confirm both the relevance and the complexity of
this picture. Small-angle scattering and microscopy have revealed finite
clusters in protein solutions and colloid--polymer mixtures
\cite{Stradner2004}, while particle-resolved confocal experiments found
anisotropic cluster growth and network formation when attraction and
electrostatic repulsion compete \cite{Campbell2005}. Tunable magnetic
dispersions provide a more recent route for controlling attraction and
repulsion in situ and directly observing the continual reconfiguration of
two-dimensional colloidal assemblies
\cite{AlHarraqBharti2022,Gauri2023}. More direct realizations of filamentous
self-assembly have been obtained with core--corona and polymer-grafted
colloids. Isotropic colloidal building blocks have been assembled into chain
phases at interfaces \cite{Rey2018,Ciarella2021}, while weakly charged
polymer-grafted particles with competing attraction and repulsion form
isolated linear and branched polymer-like clusters in aqueous suspension
\cite{Haddadi2021JCIS,Haddadi2021Langmuir}. Experiments have also produced
single-particle-wide colloidal lines \cite{HuangTao2006}, showing that
filamentary morphologies need not possess a bulk-like interior. At this
limiting thickness every particle belongs to the boundary and the local
coordination is essentially that of a chain. Consequently, bending, end
formation, reconnection, and loop closure can replace the interfacial and
bulk mechanisms that dominate the evolution of thicker lamellae.

The relation between stripes and finite particle chains is particularly
suggestive. Simulations of colloidal systems have shown that quasi-one-dimensional
aggregation may precede branching and gel formation
\cite{SciortinoTartagliaZaccarelli2005,Haw2010}, and inverse design has
demonstrated that isotropic potentials can stabilize single-stranded
``colloidomers'' as equilibrium or ergodic assemblies
\cite{BanerjeeLindquist2019}. Cluster morphology is also known to influence
whether competing-interaction fluids remain ergodic, undergo cluster-glass
arrest, or percolate into a network
\cite{SciortinoMossa2004,ManiLechner2014,
GodfrinValadez2014}. These observations suggest that a melted stripe state
should not automatically be regarded as a homogeneous disordered fluid.
Depending on the balance between end formation and reconnection, its
chain-like constituents may form finite polymer-like clusters or remain
joined into a percolated network. These alternatives can possess comparable
local packing and mobility while differing fundamentally in their global
topology.

Melting adds a second layer of complexity. For ordinary two-dimensional
crystals, the Kosterlitz--Thouless--Halperin--Nelson--Young framework relates
melting to the unbinding of topological defects
\cite{KosterlitzThouless1973,HalperinNelson1978,
NelsonHalperin1979,Young1979}. Colloidal experiments have provided direct
tests of this defect-mediated scenario and have also shown that the pathway
depends on the interaction and anisotropy
\cite{ZahnLenkeMaret1999,VonGrunbergKeim2004,Gasser2010}.
Microphase-forming systems are not ordinary crystals, however: translational
and orientational order coexist with an emergent connectivity at the domain
scale. Simulations of stripe-forming particles have accordingly found
intermediate regimes, defect-rich stripes, and nontrivial melting sequences
\cite{StoychevaSinger2000,OlsonReichhardt2010}. At the coarser level of
continuous domain patterns, the expression ``topological melting'' has
previously been used to describe the reorganization of magnetic stripes into
polygonal, bubble, and labyrinthine morphologies
\cite{Jagla2004TopologicalMelting}. That description concerns the topology of
continuous domains, however, rather than the particle-resolved connectivity,
branching, and loop structure of one-particle-thick filaments considered
here. In our previous study of the present core-softened model, the
one-particle-thick stripe phase displayed a gradual loss of order and
pronounced mobility along the stripes \cite{Puccinelli2025Melting}. What
remained unresolved was whether melting this ordered phase necessarily
destroys its system-spanning topology, and whether the resulting fluid
topology is uniquely inherited from the stripe state or selected by density.

This question cannot be answered completely by conventional scalar order
parameters. Pair correlations and bond-orientational order quantify
positional and angular organization, but they do not distinguish a
system-spanning stripe from a finite chain, a branched object, or a closed
loop when these structures have comparable local packing. Graph and
topological approaches offer the required complementary description by
encoding connected components, independent cycles, percolating or winding
objects, and cyclic backbones. More broadly, computational topology has
detected phase changes and structurally relevant length scales in statistical
models and soft materials
\cite{Donato2016,ColeLogesShiu2021,MembrilloSolis2022}, while network-based
analyses have related mesoscale connectivity to heterogeneous relaxation in
colloidal matter \cite{ColomboDelGado2014}. Nevertheless, the
temperature-resolved coupling among stripe curvature, graph topology,
filament statistics, and particle mobility has received little attention.
In particular, it remains unclear whether the same one-particle-thick stripe
phase follows a universal fragmentation pathway or can melt into
topologically distinct fluids as density changes.

Here we address this problem using molecular dynamics simulations of a
two-dimensional Lennard-Jones--Gaussian system
\cite{Oliveira06a,Bordin2023a}. We study seven densities spanning the region
in which the one-particle-thick stripe morphology is stabilized, including
the density of optimal stripe stability \cite{Cardoso2021}, and follow
continuous isochoric heating paths from identically ordered stripe
configurations. Orientational descriptors are combined with a graph
representation of the instantaneous particle structure. The first Betti
number, periodic winding, the graph 2-core, cluster- and cycle-size
distributions, filament conformations, bond persistence, long-time
self-diffusion, and connectivity-conditioned mobility resolve how bending,
opening, branching, reconnection, and fragmentation reorganize the stripe
backbones.

\section{Model and Simulation Details}
\label{sec:model}

\subsection{Interaction model and simulation protocol}
\label{sec:model_protocol}

We consider a two-dimensional monodisperse system of particles interacting
through the Lennard-Jones plus Gaussian (LJG) pair potential
\cite{Oliveira06a,BarrosdeOliveira2010b}
\begin{equation}
\begin{split}
 U(r)={}&4\varepsilon\left[
 \left(\frac{\sigma}{r}\right)^{12}
 -\left(\frac{\sigma}{r}\right)^6\right]\\
 &+A\varepsilon\exp\left[
 -\left(\frac{r-r_0\sigma}{c\sigma}\right)^2\right],
\end{split}
\label{eq:ljg_potential}
\end{equation}
where $\varepsilon$ and $\sigma$ set the energy and length scales,
respectively. We use $A=5.0$, $r_0=0.7$, and $c=1.0$, corresponding to the
shoulder-ramp core-softened model investigated in our previous studies
\cite{Cardoso2021,Puccinelli2025Melting}. The interaction was truncated at
$r_{\mathrm{cut}}=3.5\sigma$ and shifted so that
$U(r_{\mathrm{cut}})=0$. Lennard-Jones reduced units are used throughout
\cite{AllenTildesley2017}: lengths, energies, temperatures, and times are
expressed in units of $\sigma$, $\varepsilon$, $\varepsilon/k_{\mathrm B}$,
and $\tau=\sqrt{m\sigma^2/\varepsilon}$, respectively, with particle mass
$m=1$.

The molecular dynamics simulations were performed with ESPResSo
\cite{Weik2019ESPResSo}. The Lennard-Jones contribution was evaluated with
the native ESPResSo interaction, thereby preserving its short-range
divergence, while only the smooth Gaussian contribution was tabulated. The
two contributions were combined to reproduce Eq.~\eqref{eq:ljg_potential}
with the stated cutoff and shift. Periodic boundary conditions were applied
in $x$ and $y$. Motion along $z$ was constrained for every
particle and the $z$ direction was nonperiodic.

The equations of motion were integrated with the velocity-Verlet algorithm
using a time step $\Delta t=0.002\tau$. Temperature was controlled by a
Langevin thermostat with friction coefficient $\gamma=1.0\,m/\tau$. The
main simulations contained $N=4096$ particles at the seven number densities, $
 \rho=\frac{N}{L_xL_y}
=0.395,\;0.405,\;0.415,\;0.425,\;0.435,\;0.445,\;\text{and}\;0.455$.
This interval spans the region in which the one-particle-thick stripe
morphology is stabilized in the previously determined phase diagram
\cite{Cardoso2021}. 

Each run was initialized as a defect-free array of one-particle-thick
stripes. The $N=4096$ configuration contained $n_x=64$ particles along each
stripe and $n_y=N/n_x=64$ stripes. If $a_{\parallel}$ and $a_{\perp}$ denote
the particle separations parallel and perpendicular to a stripe,
respectively, we imposed
\begin{equation}
 \frac{a_{\perp}}{a_{\parallel}}=\frac{2.0}{1.2}=\frac{5}{3},
 \qquad
 a_{\parallel}=
 \left[
 \rho\left(\frac{a_{\perp}}{a_{\parallel}}\right)
 \right]^{-1/2},
\label{eq:stripe_spacings}
\end{equation}
and used $L_x=n_xa_{\parallel}$ and $L_y=n_ya_{\perp}$. This construction
places consecutive particles along a stripe near the short structural scale
and separates adjacent stripes by the longer scale. Across the simulated
density interval, $a_{\parallel}$ decreases from $1.232\sigma$ to
$1.148\sigma$, while $a_{\perp}$ decreases from $2.054\sigma$ to
$1.914\sigma$. Simultaneous matching to the reference separations
$a_{\parallel}=1.2\sigma$ and $a_{\perp}=2.0\sigma$ occurs at
$
 \rho_{\mathrm{match}}=(1.2\times2.0)^{-1}\simeq0.417$.
 The density scan therefore probes both stretched and compressed stripe
geometries relative to the two interaction-selected structural scales.

\begin{figure}[h!]
    \centering
    \includegraphics[width=\columnwidth]
      {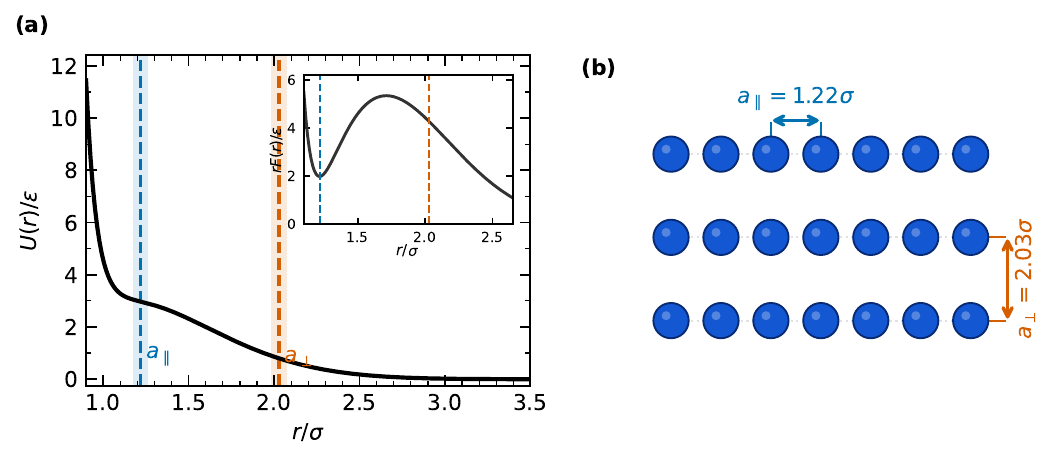}
    \caption{\label{fig:interaction_scales}
    Interaction model and microscopic organization of the stripe phase.
    (a) Lennard-Jones--Gaussian pair potential $U(r)/\varepsilon$ used in the
    simulations. The blue and orange dashed lines identify the longitudinal
    and transverse separations at $\rho=0.405$,
    $a_{\parallel}=1.22\sigma$ and $a_{\perp}=2.03\sigma$, respectively; the
    shaded regions are visual guides to these structural scales. The inset
    shows the corresponding distance-weighted radial force,
    $rF(r)/\varepsilon$. (b) Schematic representation of the
    one-particle-thick stripe arrangement. The short separation organizes
    particles along each stripe, whereas the long separation separates
    adjacent stripes.}
\end{figure}

Figure~\ref{fig:interaction_scales} summarizes the relation between the
interaction and the microscopic stripe geometry. The pair potential is
monotonic, but the distance-weighted force $rF(r)$ displays a broad softened
region that permits distinct local separations
[Fig.~\ref{fig:interaction_scales}(a)]. The marked values
$a_{\parallel}$ and $a_{\perp}$ are therefore structural scales rather than
two minima of $U(r)$. As illustrated in
Fig.~\ref{fig:interaction_scales}(b), the shorter separation organizes
consecutive particles along a one-particle-thick stripe, whereas the longer
separation separates adjacent stripes. The schematic uses $\rho=0.405$, for
which $a_{\parallel}=1.217\sigma$ and $a_{\perp}=2.029\sigma$.

Initial velocities were drawn independently from a Maxwell distribution at
$T=0.04$, and their center-of-mass contribution was removed. Four independent
realizations were simulated at each density. Continuous isochoric heating
paths were followed from $T=0.04$ to $0.20$ in increments of
$\Delta T=0.005$. The initial state point was equilibrated for $10^7$
integration steps. At every subsequent temperature, the final configuration
and velocities from the preceding state point were retained and the system
was equilibrated for $3\times10^6$ steps. Production trajectories comprised
$5\times10^7$ steps at each temperature. This protocol preserves the thermal
history of the heating trajectory without regenerating the configuration at
each state point. Particle configurations were stored every $5000$ steps,
corresponding to $\Delta t_{\mathrm{GSD}}=10\tau$, and the total energy was
sampled every $1000$ steps during production.

For each independent realization, we calculated the mean energy per particle
$e(T)=\langle E\rangle/N$. The isochoric heat capacity per particle shown in
the main text was obtained from the canonical energy-fluctuation estimator
\begin{equation}
 c_V=
 \frac{\langle E^2\rangle-\langle E\rangle^2}{NT^2},
\label{eq:cv_fluctuation}
\end{equation}
where $k_{\mathrm B}=1$ in reduced units. The numerical derivative
$\partial e/\partial T$ was used as an independent consistency check. See Supplemental Material at the end of this manuscript for
complete definitions of the observables, the robustness against the graph
connectivity cutoff, the thermodynamic response along the heating isochores,
and the system-size analysis.

System-size dependence was examined at the densities $\rho=$ 0.405, 0.425, 0.435, and 0.445,
using $N=1024$, $4096$, and $16384$. The corresponding initial configurations
contained $n_x=32$, $64$, and $128$ particles along each stripe,
respectively, with $n_y=n_x$. Thus, $\rho$,
$a_{\perp}/a_{\parallel}=5/3$, the box aspect ratio, and all interaction and
thermostat parameters were preserved as the system size changed. The resulting finite-size analysis
is presented in the Supplemental Material.

\subsection{Structural, topological, and dynamical observables}
\label{sec:observables}

The instantaneous particle configurations were represented as undirected
graphs in which particles constitute the vertices and an edge connects each
pair separated by less than
$r_{\mathrm c}^{\mathrm{graph}}=1.5\sigma$. This construction maps the
one-particle-thick stripes onto filamentous networks and permits their
connectivity, branching, fragmentation, and periodic wrapping to be followed
during heating. Robustness with respect to the graph cutoff is demonstrated
in the Supplemental Material \cite{SupplementalMaterial}.

Global and local orientational organization were characterized by the
nematic order parameter $S_2$ and the local twofold order
$\langle|\psi_2|\rangle$, while contour bending was measured by the rms
curvature. Graph topology was characterized by the first Betti number
$\beta_1$, its decomposition into contributions from components with and
without periodic winding, and the fraction of particles belonging to the
graph 2-core. Cluster-size distributions, the largest-cluster fraction, and
the percolation probability distinguish finite aggregates from
system-spanning networks. Filament geometry was further resolved through the
endpoint fraction, contour length, end-to-end distance, tortuosity, tangent
correlations, and fundamental-cycle lengths.

Fluctuations of the global orientational order were quantified by
\begin{equation}
 \chi_2=
 \frac{N}{T}
 \left(
 \langle S_2^2\rangle-\langle S_2\rangle^2
 \right),
\label{eq:orientational_susceptibility}
\end{equation}
and the temperature derivative
$\mathcal{R}_2=-\partial\langle S_2\rangle/\partial T$ was used as an
orientational response function. Together with $c_V$, these quantities locate
the structural and thermodynamic response interval without assuming a
particular order for the transition.

Particle transport was characterized by the long-time self-diffusion
coefficient obtained from the two-dimensional Einstein relation
\begin{equation}
 D_{\mathrm{MSD}}=\lim_{t\rightarrow\infty}
 \frac{\left\langle |\mathbf r_i(t)-\mathbf r_i(0)|^2\right\rangle}{4t}.
\label{eq:diffusion_msd}
\end{equation}
The coupling between connectivity and motion was probed through the
short-time mobilities of degree-one endpoints and degree-$k\geq3$ branch
particles relative to degree-two backbone particles, denoted
$R_{\mathrm{end}}$ and $R_{\mathrm{branch}}$, respectively. Bond survival
over one stored-frame interval measures the accompanying renewal of the
neighbor network.

Complete definitions of the winding criterion, graph 2-core, cluster, chain, and cycle statistics, correlation functions, mobility ratios, numerical derivatives, and uncertainty estimates are provided in the Supplemental Material \cite{SupplementalMaterial}.

\section{Results and discussion}
\label{sec:results}

\begin{figure*}[h!]
    \centering
    \includegraphics[width=\textwidth]
    {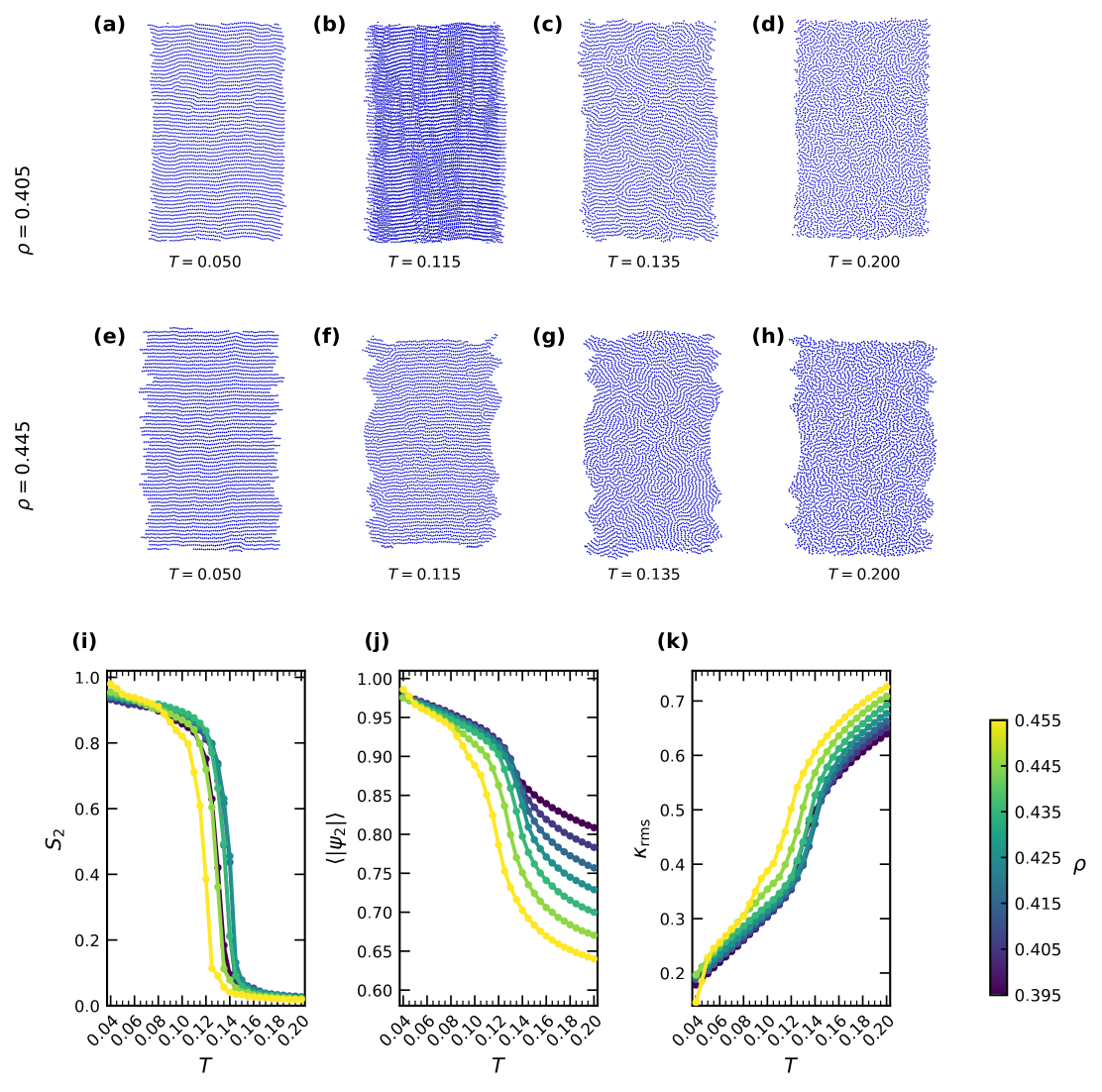}
    \caption{Density-selected pathways in the melting of one-particle-thick
    stripes. Representative configurations are shown for $\rho=0.405$ in
    (a)--(d) and $\rho=0.445$ in (e)--(h), at $T=0.050$, $0.115$, $0.135$,
    and $0.200$ from left to right. Panels (i)--(k) show, respectively, the
    global nematic order $S_2$, the local twofold order
    $\langle|\psi_2|\rangle$, and the rms contour curvature for all seven
    densities. Symbols and shaded bands denote averages and standard errors
    over four independent runs.}
    \label{fig:structural_pathway}
\end{figure*}

The thermal evolution of the one-particle-thick stripe phase is summarized in
Fig.~\ref{fig:structural_pathway}.  The representative configurations at
$\rho=0.405$ [Figs.~\ref{fig:structural_pathway}(a)--\ref{fig:structural_pathway}(d)]
and $\rho=0.445$
[Figs.~\ref{fig:structural_pathway}(e)--\ref{fig:structural_pathway}(h)] were
selected at the same temperatures to expose the two reconstruction pathways
without conflating density and thermal history.  Both systems begin as nearly
straight and mutually aligned stripes, consistent with the ordered state
previously identified for this model \cite{Puccinelli2025Melting}.  On
heating, the contours acquire
curvature and lose their common orientation.  Their final morphologies,
however, are qualitatively different: the lower-density stripes open and
fragment into finite polymer-like clusters, whereas the higher-density system
retains an interconnected structure after stripe alignment has disappeared.

The orientational observables show that the initial loss of stripe order is
nevertheless similar in the two pathways.  The global nematic order $S_2$
decreases sharply over a narrow temperature interval
[Fig.~\ref{fig:structural_pathway}(i)], while the local twofold order
$\langle|\psi_2|\rangle$ remains comparatively high
[Fig.~\ref{fig:structural_pathway}(j)].  The midpoint temperature of the
$S_2$ decay first increases from approximately $0.128$ at $\rho=0.395$ to
$0.139$ at $\rho=0.415$--$0.425$, and then decreases to $0.117$ at
$\rho=0.455$.  Thus, the extended density range reveals a nonmonotonic
stability of global stripe alignment rather than the continuous upward shift
suggested by the three central densities alone.  At these midpoints,
$\langle|\psi_2|\rangle$ remains between approximately $0.81$ and $0.90$.
Local anisotropic environments therefore survive the decay of global
alignment throughout the density interval.

The simultaneous increase of the rms curvature
[Fig.~\ref{fig:structural_pathway}(k)] identifies bending as the geometrical
precursor of this loss of alignment.  Because the stripes are only one
particle thick, they have no bulk-like interior that can disorder
independently of their boundaries.  Thermal fluctuations act directly on
their contours, making bending, end formation, reconnection, and
fragmentation the relevant relaxation modes.  This differs from finite-width
stripes, for which internal disordering and intrastripe motion can precede
large-scale breakup \cite{OlsonReichhardt2010}.  Figures
\ref{fig:structural_pathway}(i)--\ref{fig:structural_pathway}(k) consequently
establish a common orientational melting process, but the configurations in
Figs.~\ref{fig:structural_pathway}(d) and
\ref{fig:structural_pathway}(h) already indicate that its topological outcome
depends strongly on density.

\begin{figure*}[h!]
    \centering
    \includegraphics[width=\textwidth]
    {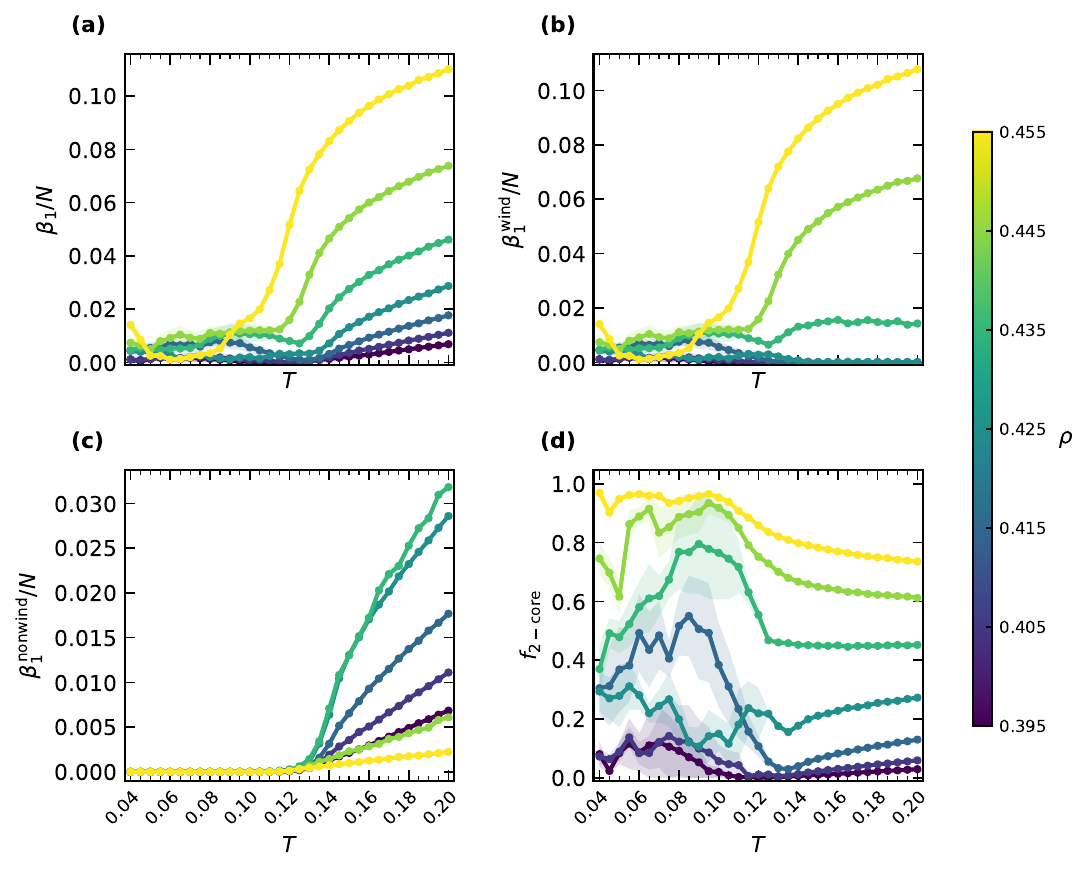}
    \caption{Topological reconstruction during heating for the seven
    densities. (a) First Betti number per particle, $\beta_1/N$.
    Contributions to the cycle rank from components (b) with and (c) without
    nonzero periodic winding; cycles counted in (b) need not individually
    wind around the box. (d) Fraction of particles in the graph two-core. Symbols and
    shaded bands denote averages and standard errors over four independent
    runs.}
    \label{fig:topological_reconstruction}
\end{figure*}

The connectivity observables in
Fig.~\ref{fig:topological_reconstruction} clarifies this point.  Most
importantly, the nonmonotonicity of $\beta_1$ is itself density dependent:
the curves do not represent a common sequence translated along the
temperature axis.  For $\rho=0.395$--$0.415$, the total cycle density first
displays a low-temperature shoulder, then decreases to a pronounced minimum,
and finally increases again
[Fig.~\ref{fig:topological_reconstruction}(a)].  The minima occur near
$T=0.110$, $0.115$, and $0.130$, respectively.  At the two highest
densities, by contrast, the dominant feature is the strong increase of
$\beta_1/N$ through and beyond the loss of orientational order.  The
intermediate densities connect these limits in a nontrivial way.

The winding decomposition is essential for identifying the origin of these
different curve shapes.  At low temperature, essentially the entire
$\beta_1$ is carried by components with nonzero periodic winding
[Fig.~\ref{fig:topological_reconstruction}(b)], while the nonwinding
contribution remains negligible
[Fig.~\ref{fig:topological_reconstruction}(c)].  The initial shoulder at the
three lower densities therefore does not originate from compact loops in a
cluster fluid.  It reflects opening, closure, and reconnection within the
system-spanning stripe pattern.  Since every stripe is only one particle
thick, a local reconnection can immediately change the cycle rank of the
spanning component even while $S_2$ and $\langle|\psi_2|\rangle$ remain
large.  The ordered stripe state is consequently not topologically inert.

For $\rho=0.395$--$0.425$, further heating removes the winding contribution
and replaces it with cycles belonging to finite, nonwinding objects.  The
two contributions become equal at
$T_{\rm cross}\simeq0.115$, $0.123$, $0.128$, and $0.132$, respectively.
Around this crossover, destruction of the original winding backbones is not
yet compensated by the proliferation of local loops, producing the minimum
or weak-cycle interval in the total $\beta_1$.  The structures are therefore
predominantly open and curved: they preserve chain-like local packing but
have lost most of the global connectivity of the stripe state.  We use
\emph{topological intermediate} for this connectivity regime without
implying a separate thermodynamic phase.  Its finite temperature width shows
that opening a winding stripe and closing new loops among its fragments are
distinct processes.

The evolution at $\rho=0.435$ is qualitatively mixed.  The winding
contribution decreases from $\beta_1^{\rm wind}/N\simeq0.0106$ at $T=0.10$
to approximately $0.0065$ at $T=0.125$, but then increases again to
approximately $0.0156$ at $T=0.16$.  Nonwinding cycles grow simultaneously
and become dominant only at $T_{\rm cross}\simeq0.155$, after the main decay
of $S_2$.  Thus, this density does not merely postpone fragmentation.  It
samples a competition between the production of finite loops and the
reconstruction of cycles inside components that remain system spanning.

This second mechanism becomes dominant at $\rho=0.445$ and $0.455$.  For
$\rho=0.445$, $\beta_1^{\rm wind}/N$ grows from approximately $0.012$ at
$T=0.10$ to $0.068$ at $T=0.20$; for $\rho=0.455$, it grows from
approximately $0.0165$ to $0.108$ over the same interval.  Much of this
growth occurs after the maxima of the orientational response, when the
straight stripe alignment has already disappeared.  No winding--nonwinding
crossover is reached by $T=0.20$, and the corresponding nonwinding
contributions are only about $0.0061$ and $0.0022$.  Heating therefore does
not simply preserve a progressively weakened remnant of the initial stripe
pattern.  It produces additional independent cycles within a winding,
orientationally disordered network.

Here $\beta_1^{\rm wind}$ denotes the cycle rank of every connected component
that has nonzero periodic winding; an individual cycle contributing to this
quantity need not itself traverse the box.  Its post-melting growth can thus
be generated when branch formation and reconnection add short local loops to
an already winding backbone.  This distinction reconciles the rapid increase
of $\beta_1^{\rm wind}$ with the coexistence of short and system-scale cycles
shown below: the high-density fluid is simultaneously locally reticulated and
globally connected.

The two-core fraction in Fig.~\ref{fig:topological_reconstruction}(d)
independently measures how much of the structure belongs to a cyclic
backbone.  At $T=0.20$, it increases from approximately $0.029$ at
$\rho=0.395$ to $0.273$ at $\rho=0.425$, $0.452$ at $\rho=0.435$, and
$0.612$ and $0.736$ at $\rho=0.445$ and $0.455$.  The low-density increase
of $\beta_1$ is therefore produced by numerous small loops involving a
limited particle fraction, whereas the high-density increase is accompanied
by an extensive cyclic backbone.  The density dependence of $\beta_1$ thus
separates two physically different outcomes of melting: finite loop-bearing
filaments and a multiply connected winding fluid.

This interpretation extends defect-based descriptions of stripe melting
\cite{StoychevaSinger2000} by distinguishing the opening of winding
backbones, the creation of local cycles, and the reticulation of a spanning
component.  More generally, experiments on two-dimensional colloidal
melting demonstrate that the production and evolution of topological
defects need not coincide with a single loss-of-order temperature
\cite{Vyas2026}.

\begin{figure*}[h!]
    \centering
    \includegraphics[width=\textwidth]
    {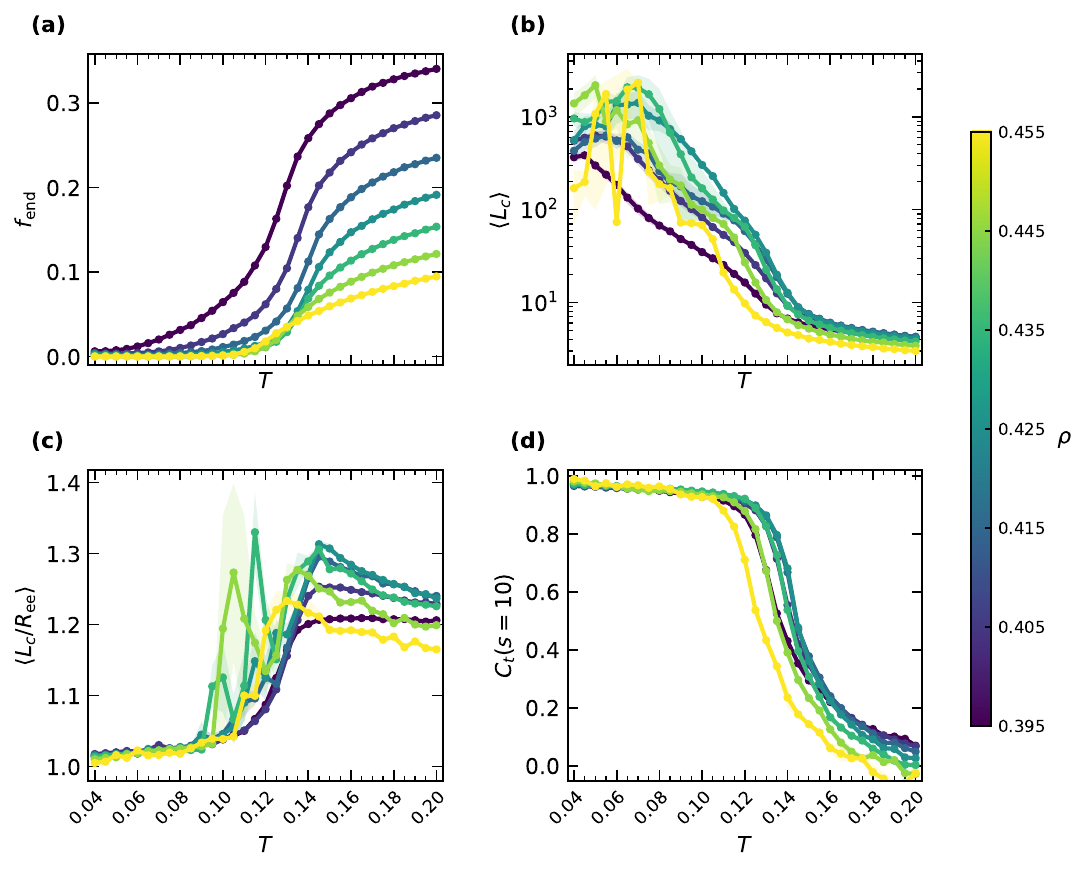}
    \caption{Evolution of filament geometry during heating.
    (a) Endpoint fraction, (b) mean contour length, (c) mean tortuosity
    $\langle L_c/R_{\rm ee}\rangle$, and (d) tangent correlation at contour
    separation $s=10$. Symbols and shaded bands denote averages and standard
    errors over four independent runs.}
    \label{fig:chain_geometry}
\end{figure*}

The filament statistics in Fig.~\ref{fig:chain_geometry} connect this
topological distinction to the geometry of the aggregates.  The endpoint
fraction increases throughout the reconstruction
[Fig.~\ref{fig:chain_geometry}(a)], but its final value decreases markedly
with density, from approximately $0.34$ at $\rho=0.395$ to $0.19$ at
$\rho=0.425$ and $0.095$ at $\rho=0.455$.  End formation is therefore the
dominant outcome on the low-density side, whereas it is progressively
suppressed as the connected network becomes favored.  The characteristic
contour length simultaneously decreases by orders of magnitude
[Fig.~\ref{fig:chain_geometry}(b)], showing that loss of system-spanning
stripe order produces shorter connected segments in both regimes.

The opposite density trends of the endpoint and two-core fractions expose
the local operations responsible for the two topological outcomes.  On the
low-density side, opening a winding contour removes particles from the
cyclic backbone and creates two mobile ends; the large endpoint fraction and
small two-core fraction at $T=0.20$ are the cumulative signatures of this
process.  On the high-density side, contacts between bent filaments create
junctions and alternative paths.  End production is then suppressed while
the two-core expands, and the newly formed loops remain attached to a
winding component.  The change between these limits is therefore not merely
a variation in cluster size: density selects whether bond rearrangements
terminate a filament or reticulate the network.

The conformational evolution is not a direct collapse into compact isotropic
aggregates.  The mean tortuosity rises from values close to unity to
approximately $1.2$--$1.3$ [Fig.~\ref{fig:chain_geometry}(c)], while the
tangent correlation at contour separation $s=10$ loses its initial
long-range coherence [Fig.~\ref{fig:chain_geometry}(d)].  Near the
topological intermediate, the contours are already substantially curved
although many remain much longer than the final clusters.  The subsequent
weak decrease of tortuosity at the highest temperatures does not represent a
recovery of straight stripes; it follows from the shortening of the
available contour over which large-scale bending can develop.

At lower density, the loss of tangent coherence accompanies the conversion
of long open filaments into flexible finite polymer-like clusters.  At
higher density, the same local loss of coherence occurs without global
fragmentation: the measured contours are flexible strands connecting
junctions inside a multiply connected network.  Hence similar local chain
statistics can coexist with opposite global topologies.  Linear growth
followed by branching is a known route to network formation in colloids with
competing interactions \cite{SciortinoTartagliaZaccarelli2005}, and both
linear and branched aggregates have been observed in experiments and
particle-based models
\cite{Haddadi2021JCIS,Haddadi2021Langmuir}.  The present results show that
the balance between these motifs is continuously selected by density during
the melting of an initially ordered stripe phase.

\begin{figure*}[h!]
    \centering
    \includegraphics[width=\textwidth]
    {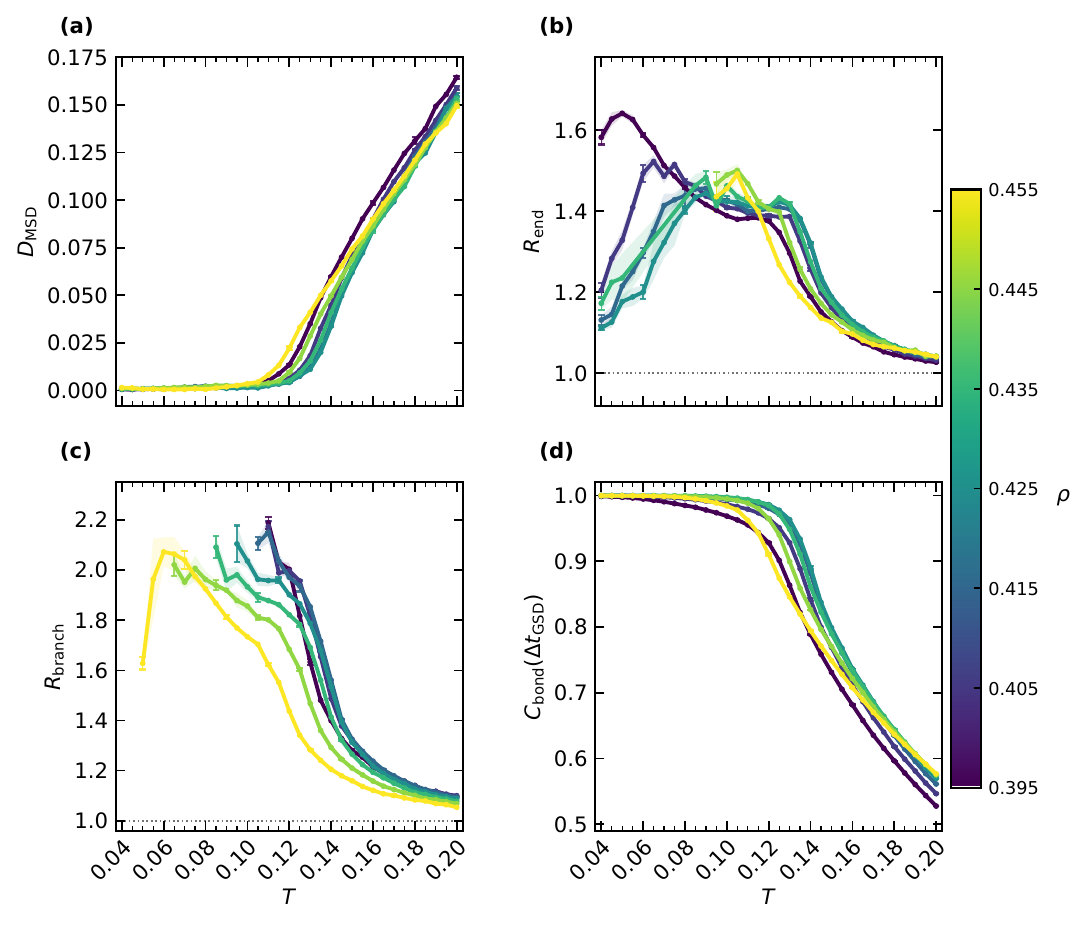}
    \caption{Coupling between local connectivity and dynamics. (a) Long-time diffusion
coefficient $D_{\rm MSD}$, (b) mobility of filament ends relative to the
degree-two backbone, $R_{\rm end}$, (c) relative mobility of branch
particles, $R_{\rm branch}$, and (d) bond-survival probability over one
stored-frame interval. Topology-conditioned mobility ratios are reported
only when all four replicas contain, on average, at least two particles of
the corresponding class per stored configuration. Shaded bands and sparse
error bars denote standard errors over independent replicas.}
    \label{fig:topology_dynamics}
\end{figure*}

The two final morphologies are dynamically fluid.  The long-time diffusion
coefficient increases smoothly through the reconstruction interval
[Fig.~\ref{fig:topology_dynamics}(a)] and remains only weakly density
dependent.  At $T=0.20$, $D_{\rm MSD}$ decreases modestly from
$0.1645\pm0.0009$ at $\rho=0.395$ to $0.1496\pm0.0010$ at $\rho=0.455$.
In particular, the percolated structures at $\rho=0.445$ and $0.455$ have
diffusion coefficients comparable to those of the finite-cluster fluids.
Their persistent connectivity should therefore not be interpreted as
network arrest or gelation; it characterizes the topology of a continuously
restructuring fluid.  It is relevant because competing
interactions can also produce arrested cluster and network states
\cite{SciortinoMossa2004}; here, the finite diffusivity demonstrates that
percolation and arrest are separated.

Conditioning the displacement on local graph environment reveals dynamical
heterogeneity that is hidden in the bulk diffusivity.  Filament ends are more
mobile than degree-two backbone particles
[Fig.~\ref{fig:topology_dynamics}(b)].  Their relative mobility
$R_{\rm end}$ first increases, then forms a broad plateau near
$1.3$--$1.5$, and finally approaches unity.  The initial enhancement shows
that terminations are mobile defects of the stripe pattern.  At the lower
densities, this enhancement precedes and overlaps the collapse of
$\beta_1^{\rm wind}$: opening a spanning contour simultaneously creates an
end and releases a locally undercoordinated particle.  As endpoints
proliferate and the degree-two segments themselves become mobile, the end
environment becomes less exceptional, accounting for the plateau and
subsequent decay of the mobility contrast.  At the highest densities, the endpoint population is negligible in the
low-temperature stripe phase. We therefore report $R_{\rm end}$ only when
all four replicas contain, on average, at least one complete endpoint pair
per stored configuration. This population criterion removes ratios formed
from an effectively absent conditioning class and does not affect the
reconstruction interval.

Branch particles display a distinct temporal hierarchy
[Fig.~\ref{fig:topology_dynamics}(c)].  On the low-density side, a statistically
resolved branch population appears only after fragmentation has begun, with
an initial mobility close to twice that of the backbone.  Increasing density
makes junctions abundant at lower temperature and simultaneously suppresses
the production of endpoints.  Branches are consequently not the direct
product of bond breaking: they arise when bent contours contact, merge, or
reconnect, and become the elements that maintain connectivity in the dense
network fluid.  Their initial mobility, approximately $1.6$--$2$ times that
of the backbone, identifies junctions as active reconstruction sites rather
than permanent crosslinks.  At $\rho=0.445$ and $0.455$, the interval with a
well-developed branch signal is also the interval in which
$\beta_1^{\rm wind}$ grows rapidly.  The structural and dynamical evidence
therefore support the same mechanism: mobile contacts create alternative
paths and local loops while retaining the system-spanning component.  Their
mobility contrast, like that of the endpoints, decays toward unity at high
temperature, when bond rearrangements are no longer localized at a special
coordination environment.

The bond-survival probability decreases from nearly one to values between
approximately $0.53$ and $0.58$
[Fig.~\ref{fig:topology_dynamics}(d)].  Its strongest variation overlaps the
loss of tangent coherence and the redistribution between winding and
nonwinding cycles.  The persistent high-density network is therefore not a
fixed set of bonds.  It maintains global connectivity while its individual
connections are continually renewed, consistent with the comparable
diffusivities in Fig.~\ref{fig:topology_dynamics}(a).  At low density this
renewal primarily advances fragmentation; at high density it changes the
internal cycle structure of the winding component without destroying its
global connectivity.  Similar connections
between mesoscale connectivity and heterogeneous dynamics have been reported
for colloidal networks and competing-interaction fluids
\cite{ColomboDelGado2014,Gauri2023}.

\begin{figure*}[h!]
    \centering
    \includegraphics[width=\textwidth]
    {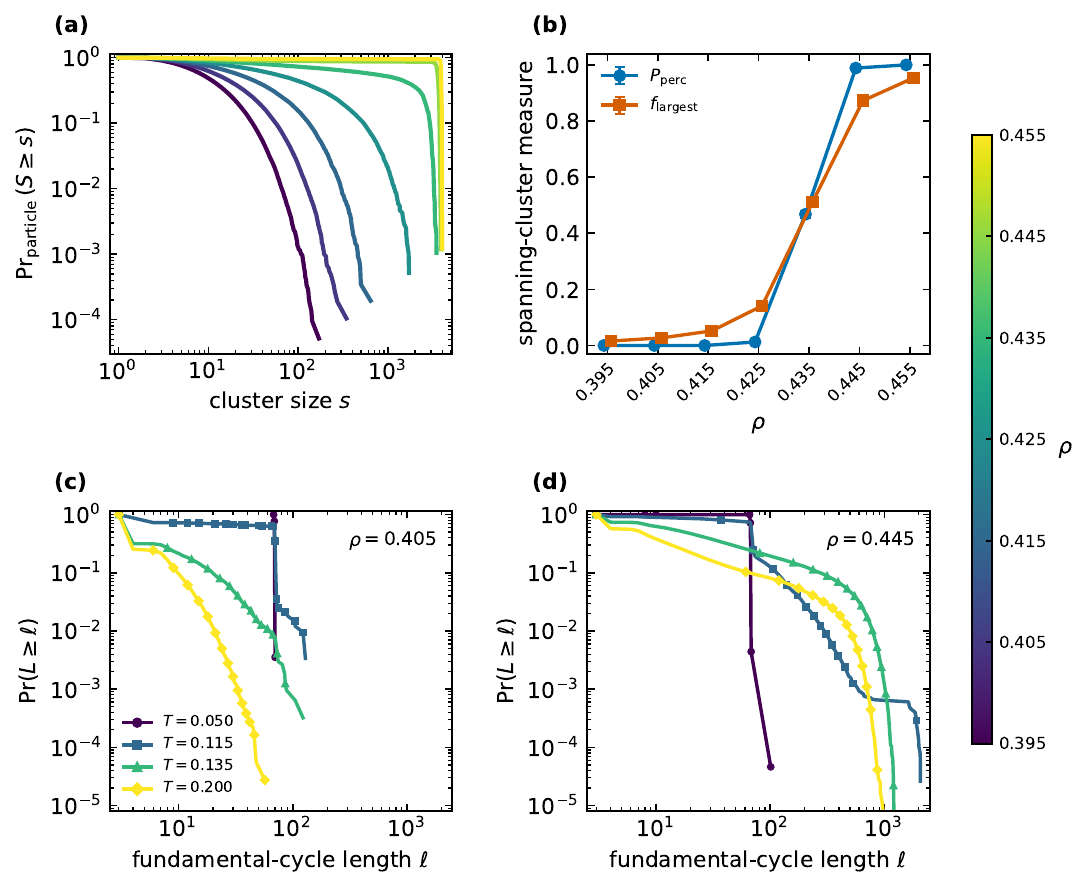}
    \caption{Density-selected topology of the high-temperature fluid.
    (a) Particle-weighted cluster-size complementary cumulative
    distributions at $T=0.20$. (b) Percolation probability and fraction of
    particles in the largest cluster at the same temperature.
    Number-weighted complementary cumulative distributions of fundamental
    cycle lengths at selected temperatures for (c) $\rho=0.405$ and
    (d) $\rho=0.445$. Error bars in (b) are standard errors over four
    independent runs and are smaller than the data points.}
    \label{fig:density_selected_topology}
\end{figure*}

The distributions in Fig.~\ref{fig:density_selected_topology} establish the
length scales and spanning character of the two fluids.  The
particle-weighted cluster-size distributions at $T=0.20$
[Fig.~\ref{fig:density_selected_topology}(a)] shift continuously toward larger
objects as density increases.  For $\rho\leq0.425$, the distributions decay
before the system size, confirming a fluid of finite clusters.  At
$\rho=0.445$ and $0.455$, they instead retain most of their weight near the
system size, consistent with a dominant connected network.  The strong
density dependence is consistent with the general sensitivity of cluster
morphology and percolation to thermodynamic conditions in
competing-interaction fluids
\cite{ValadezPerez2021,ZhuangZhangCharbonneau2016}.

The spanning-cluster measures in
Fig.~\ref{fig:density_selected_topology}(b) make this crossover explicit.
At $\rho=0.425$, the percolation probability is only $0.012$ and the largest
cluster contains approximately $14\%$ of the particles.  At $\rho=0.435$,
these quantities increase to $0.468$ and $0.511$, respectively, identifying a
crossover regime in which finite and spanning structures are both sampled.
At $\rho=0.445$, the percolation probability is already $0.989$ and the
largest component contains about $87\%$ of the particles; at $\rho=0.455$,
the corresponding values are $1.0$ and $95\%$.  Because percolation is
sensitive to finite size and to the connectivity cutoff, these results do not
by themselves locate a sharp density-driven transition.  They robustly
distinguish, however, the finite-cluster and percolated-network limits on
opposite sides of the crossover.

The cycle-size distributions further show that the distinction is not
captured solely by the presence of one large cluster.  At $\rho=0.405$
[Fig.~\ref{fig:density_selected_topology}(c)], the low-temperature
distribution is concentrated near the stripe winding length.  This feature
disappears during heating, and by $T=0.20$ the median fundamental-cycle length
is three particles, with a $90$th percentile of approximately nine.  The
renewed high-temperature growth of $\beta_1$ at this density therefore
originates from short local loops in finite aggregates.

At $\rho=0.445$ [Fig.~\ref{fig:density_selected_topology}(d)], short cycles
also become numerous after loss of stripe order, but they coexist with a much
broader tail.  At $T=0.20$, the median cycle length is approximately seven,
the $90$th percentile is about $65$, and rare cycles extend to more than
$10^3$ particles.  Local restructuring and system-spanning connectivity
therefore coexist in the dense fluid.  Because short loops attached to the
percolated backbone contribute to $\beta_1^{\rm wind}$, this broad,
two-scale distribution provides the geometrical counterpart of its
post-melting increase in Fig.~\ref{fig:topological_reconstruction}(b).
Branching points can mediate
restructuring between locally ordered aggregate segments
\cite{Opdam2025}; in the present system, increasing density converts this
local reconnection mechanism into the means by which a winding network
survives orientational melting.

\begin{figure*}[h!]
    \centering
    \includegraphics[width=\textwidth]
    {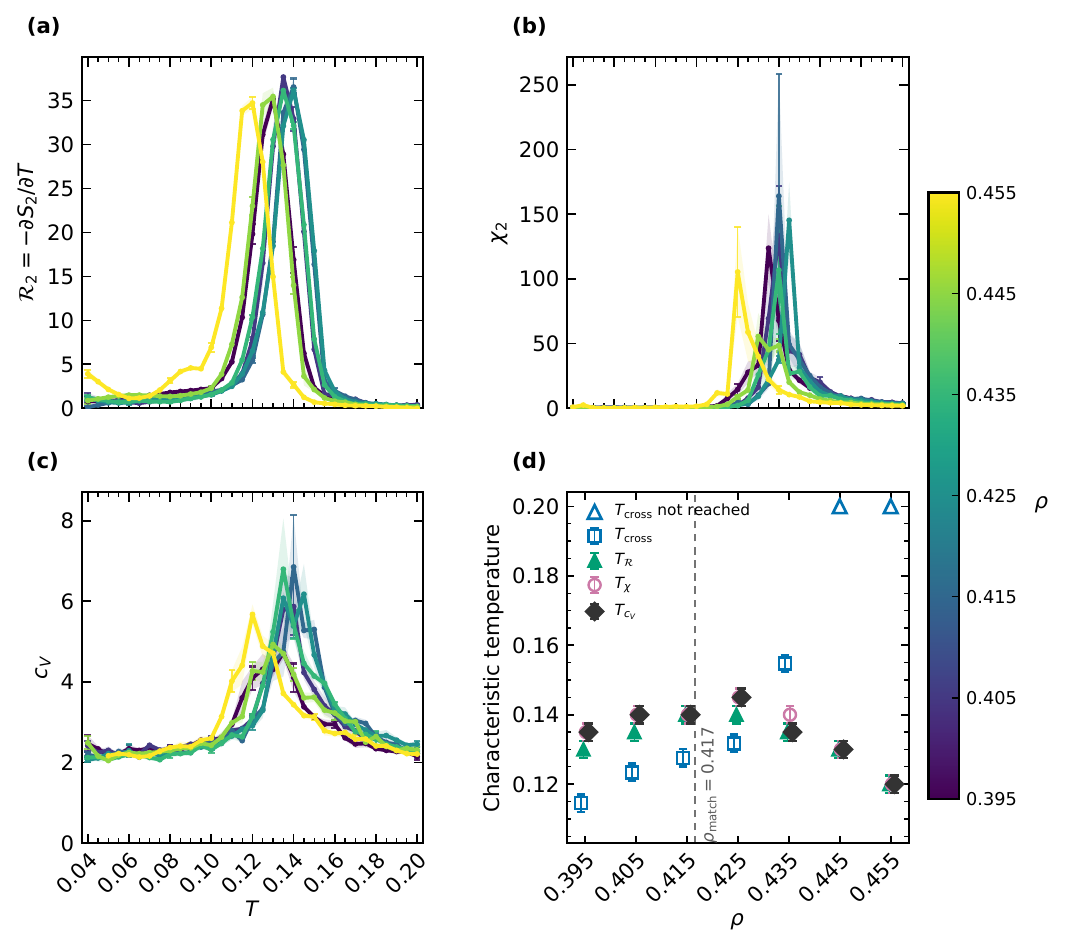}
    \caption{Structural and thermodynamic response across density. (a) Orientational
response $\mathcal{R}_2=-\partial S_2/\partial T$, (b) orientational
susceptibility $\chi_2$, and (c) heat capacity per particle $c_V$. The two
lowest-temperature state points at $\rho=0.455$ are omitted from panel (c)
because the energy records retain nonstationary relaxation of the prepared
initial configuration; they are not thermodynamic fluctuation estimates.
(d) Density dependence of the winding--nonwinding crossover temperature
$T_{\rm cross}$ and the maxima of the three response functions. The vertical
dashed line marks $\rho_{\rm match}=(1.2\times2.0)^{-1}\simeq0.417$.
Open triangles at $T=0.20$ identify densities for which $T_{\rm cross}$ was
not reached. Shaded bands and sparse error bars in (a)--(c) denote standard
errors over independent replicas; temperature error bars in (d) represent
half the simulated temperature increment.}
    \label{fig:density_response_map}
\end{figure*}

The relation between the structural, topological, and thermodynamic
temperature scales is collected in Fig.~\ref{fig:density_response_map}.  The
orientational response $\mathcal{R}_2$ has a well-defined maximum at every
density [Fig.~\ref{fig:density_response_map}(a)].  The susceptibility
$\chi_2$ [Fig.~\ref{fig:density_response_map}(b)] and the heat capacity
$c_V$ [Fig.~\ref{fig:density_response_map}(c)] peak in the same temperature
interval within the resolution $\Delta T=0.005$.  Their common density
dependence is nonmonotonic: the characteristic temperature increases from
approximately $0.13$--$0.135$ at $\rho=0.395$ to $0.14$--$0.145$ around
$\rho=0.415$--$0.425$, and then decreases to $0.12$ at $\rho=0.455$.

The characteristic-temperature map in
Fig.~\ref{fig:density_response_map}(d) demonstrates that coincidence of the
orientational and thermodynamic responses does not imply a simultaneous
topological conversion.  Up to $\rho=0.425$, $T_{\rm cross}$ rises toward the
response line while remaining slightly lower.  At $\rho=0.435$, it instead
moves to approximately $0.155$, above the maxima of
$\mathcal{R}_2$, $\chi_2$, and $c_V$.  At $\rho=0.445$ and $0.455$, a
winding-to-nonwinding crossover is not reached within the simulated
temperature interval, as indicated by the open triangles.  The thermodynamic
signature is thus associated primarily with loss of stripe orientational
order, whereas the topology of the disordered fluid is selected by density
and can remain winding well beyond that loss.

The nonmonotonic response line can be related to the two geometrical scales
of the stripe construction.  With
$a_{\perp}/a_{\parallel}=5/3$, simultaneous matching to
$a_{\parallel}=1.2\sigma$ and $a_{\perp}=2.0\sigma$ occurs at
$\rho_{\rm match}=(1.2\times2.0)^{-1}\simeq0.417$, indicated by the vertical
dashed line in Fig.~\ref{fig:density_response_map}(d).  The maximum of the
orientational and thermodynamic response temperatures lies close to this
reference.  Moving to either side stretches or compresses both imposed
separations relative to these interaction-selected values, providing a
natural source of geometric frustration. This geometrical matching is consistent with the observed nonmonotonic
density dependence of stripe stability, although it should be regarded as
a structural reference rather than an independent determination of a phase
boundary.  The conversion to the connected
high-density fluid occurs only beyond this matching point.  For example,
$a_{\perp}$ decreases from approximately $2.00\sigma$ at $\rho=0.415$ to
$1.94\sigma$ at $\rho=0.445$, increasing the opportunity for contacts and
reconnections between neighboring filaments.  Such sensitivity of modulated
morphologies to competing scales and geometric frustration is a generic
feature of stripe-forming systems
\cite{SeulAndelman1995,CiachPekalskiGozdz2013}.

At fixed $N$, increasing density also reduces both periodic box lengths as
$\rho^{-1/2}$; between $\rho=0.395$ and $0.455$ the reduction is about
$6.8\%$.  This can enhance the probability that a connected object closes
through a periodic boundary and therefore contributes to the measured
winding and percolation probabilities.  It is not, however, equivalent to
reducing the number of particles along a stripe, which remains fixed in the
present construction.  We consequently interpret the interval
$\rho=0.435$--$0.445$ as a finite-size-broadened connectivity crossover between cluster- and
network-dominated fluids rather than assign a sharp density-driven
transition to it.

The system-size analysis reported in the Supplemental Material shows that
the high-density network is not generated solely by periodic closure in a
small simulation box. At $\rho=0.445$ and $T=0.18$, the winding probability
remains above $0.979$, while the largest connected component contains
approximately $89\%$ of the particles for $N=1024$, $4096$, and $16384$.
Moreover, $\beta_1/N$ and the 2-core fraction closely collapse for the
three sizes. In contrast, the winding probability and largest-cluster
fraction decrease with $N$ on the low-density side. These findings support
a robust distinction between the finite-cluster and winding-network
regimes, while confirming that the intermediate density range remains a
finite-size-broadened connectivity crossover.

\section{Conclusion}
\label{sec:conclu}

The thermal disordering of one-particle-thick stripes proceeds through
coupled but distinct structural, topological, and dynamical changes. Upon
heating, the initially aligned stripes bend and lose global orientational
coherence while retaining substantial local twofold organization. The
orientational response, susceptibility, and heat capacity identify a common
thermal interval for this loss of stripe order, whereas connectivity evolves
over a broader and density-dependent temperature range. Stripe melting is
therefore not a single transformation in which all signatures of order
disappear simultaneously.

Density determines the topology selected after orientational melting. At
lower densities, winding stripe backbones open and fragment into finite
polymer-like clusters containing predominantly short, nonwinding loops. At
higher densities, branching and reconnection preserve a cycle-rich,
system-spanning network, with intermediate densities displaying a crossover
between these outcomes. This distinction also explains the density-dependent
nonmonotonicity of $\beta_1$: its high-temperature increase reflects local
cyclization within finite clusters at low density, but reticulation of a
winding network at high density, rather than recovery of the ordered stripe
topology. Both regimes remain diffusive, so the persistent high-density
network is a continuously restructuring fluid rather than an arrested gel.

Topology-conditioned mobilities provide a microscopic picture of this
reconstruction. Endpoints become mobile before the principal orientational
transformation, whereas branches emerge later through contacts and
reconnections between curved filament segments. This close coupling between
contour fluctuations and connectivity follows from the one-particle
thickness of the stripes, which leaves no bulk-like interior that can
disorder independently of their boundaries. More broadly, the results show
that orientational and thermodynamic observables alone do not determine the
connectivity of the melted state. Combining them with graph topology and
particle dynamics provides a general framework for describing the melting of
narrow filamentary microphases.

\begin{acknowledgments}

Without public funding, this research would be impossible. The author acknowledges financial support from Brazilian National Council for Scientific and Technological Development (CNPq), grant n$^o$ 309600/2026-0, from Fundação de Amparo à Pesquisa do Estado do Rio Grande do Sul (FAPERGS), grant n$^o$ 25/2551-0002609-1, and from the  Coordination for the Improvement of Higher Education Personnel (CAPES - Finance Code 001) and the Alexander von Humboldt Foundation for financial support through a research fellowship at the University of Konstanz. 
\end{acknowledgments}

\section*{Data Availability Statement}

The data that support the findings of this study were generated by numerical simulations. The source code and parameters used to generate the simulations and to analyze the trajectories are publicly available in~\cite{zenodofiles}.

\newpage

\begin{center}
\huge{Supplemental Material}

\end{center}

\
\section*{S1. Details on structural, topological, and dynamical observables}
\label{sec:SMmodel_observables}

The instantaneous particle configuration was represented by an undirected
geometric graph $G=(V,E)$. Each particle defines a vertex and two particles
are joined by an edge when their minimum-image separation is smaller than
$r_{\mathrm b}=1.5\sigma$. This distance lies between the two characteristic
structural scales of the model and separates particles belonging to the same
filament from particles in neighboring stripes. The robustness of the
topological results was examined by repeating the graph analysis for
$1.35\sigma\leq r_{\mathrm b}\leq1.65\sigma$ in increments of
$0.05\sigma$. The resulting cutoff maps are reported in Fig.~S1; the
nonmonotonic evolution
of $\beta_1$ and the density-dependent position of its minimum persist
throughout the physically relevant first-neighbor interval.

The global stripe orientation was quantified by the nematic order parameter
\begin{equation}
 S_2=\left|\frac{1}{N_{\mathrm b}}
 \sum_{(i,j)\in E}\exp(2\mathrm{i}\theta_{ij})\right|,
 \label{eq:S2}
\end{equation}
where $N_{\mathrm b}=|E|$ and $\theta_{ij}$ is the polar angle of the
minimum-image bond vector joining particles $i$ and $j$. The local anisotropy of the neighbor environment was quantified by~\cite{Steinhardt1983}
\begin{equation}
 \psi_{2}^{\mathrm{loc}}=
 \frac{1}{N^\prime}\sum_{i:k_i>0}
 \left|\frac{1}{k_i}\sum_{j\in{\cal N}(i)}
 \exp(2\mathrm{i}\theta_{ij})\right|,
 \label{eq:psi2}
\end{equation}
where ${\cal N}(i)$ and $k_i$ are the neighbor set and degree of particle
$i$, respectively, and $N^\prime$ is the number of nonisolated particles.
We refer to this quantity as the local twofold bond anisotropy. In
particular, it should not be interpreted independently as a local-order
parameter because its magnitude can remain finite in low-coordination
environments, including filament endpoints.

The spatial coherence of the local directors was therefore characterized
through
\begin{equation}
 C_2(r)=
 \left\langle
 \mathrm{Re}\left[q_2(i)q_2^*(j)\right]
 \right\rangle_{|\mathbf r_i-\mathbf r_j|=r},
 \label{eq:C2}
\end{equation}
where
\begin{equation}
 q_2(i)=
 \frac{1}{k_i}\sum_{j\in{\cal N}(i)}
 \exp(2\mathrm{i}\theta_{ij}).
\end{equation}
While $S_2$ measures alignment over the entire system,
$\psi_{2}^{\mathrm{loc}}$ characterizes the instantaneous anisotropy of
individual neighbor environments and $C_2(r)$ determines the spatial range
over which their orientations remain correlated.

Local bending was evaluated at vertices of degree two. For the two outward
bond vectors $\mathbf u$ and $\mathbf v$, the turning angle was defined as
\begin{equation}
 \vartheta_i=\cos^{-1}\left(
 -\frac{\mathbf u\cdot\mathbf v}{|\mathbf u||\mathbf v|}
 \right),
\end{equation}
so that $\vartheta_i=0$ for a locally straight filament. The corresponding
discrete curvature is
\begin{equation}
 \kappa_i=\frac{\vartheta_i}{(|\mathbf u|+|\mathbf v|)/2},
 \qquad
 \kappa_{\mathrm{rms}}=
 \sqrt{\left\langle\kappa_i^2\right\rangle}.
 \label{eq:curvature}
\end{equation}
Particle degrees also provide a local classification of the graph:
$k_i=1$ identifies an endpoint, $k_i=2$ a filament backbone, and
$k_i\geq3$ a branch or junction.

Connected components of $G$ were identified under periodic boundary
conditions. Periodic-image offsets were propagated along graph edges, which
allowed components with nonzero winding in either the $x$ or $y$ direction
to be distinguished from finite, nonwinding clusters. If $C$ is the number
of connected components, the number of independent cycles is
\begin{equation}
 \beta_1=|E|-|V|+C.
 \label{eq:betti}
\end{equation}
For each connected component $c$, we also evaluated
$\beta_1^{(c)}=E_c-V_c+1$ and separated the total cycle rank into
contributions from winding and nonwinding components. This distinction is
essential because a box-spanning stripe and a finite local loop both
contribute to $\beta_1$, despite representing different physical
structures.
The use of graph and homological descriptors to resolve structural motifs in
particle configurations is consistent with their established application to
condensed and disordered matter \cite{Hiraoka2016,Wang2025GraphDescriptors}.
The explicit winding classification is the periodic-system analogue of the
wrapping criteria used to distinguish finite connectivity from percolation
on a torus \cite{MertensZiff2016}.

A configuration was classified as percolating when at least one connected
component had nonzero winding under the periodic boundary conditions. The
fraction of analyzed configurations satisfying this criterion is denoted by
$P_{\rm wind}$.

The cyclic backbone was further characterized by the graph 2-core, obtained
by recursively removing all vertices with degree smaller than two. We
recorded the fraction of particles in the 2-core and the size distribution
of its connected components. The $k$-core decomposition is a standard means
of isolating the mutually supported interior of a network
\cite{Dorogovtsev2006}. Fundamental-cycle lengths were obtained from a
Paton cycle basis \cite{Paton1969}. The number of elements in this basis is exactly
$\beta_1$, although the length assigned to an individual basis cycle is not
a topological invariant and can depend on the selected basis. These lengths
are therefore used as a geometrical measure that distinguishes stripe-scale
cycles from short local loops.

Filament statistics were restricted to finite, nonwinding, unbranched, open
components: each accepted component has exactly two endpoints, no vertex of
degree larger than two, and no cycle. For an ordered sequence of bond vectors
$\{\mathbf b_n\}$ along such a filament, the contour and end-to-end lengths
are
\begin{equation}
 L_{\mathrm c}=\sum_n|\mathbf b_n|,
 \qquad
 R_{\mathrm{ee}}=\left|\sum_n\mathbf b_n\right|,
 \label{eq:chain_lengths}
\end{equation}
where minimum-image bond vectors were accumulated to unwrap the chain. The
tortuosity is $\mathcal{T}=L_{\mathrm c}/R_{\mathrm{ee}}$. Orientational
memory along a contour was measured through the tangent correlation
\begin{equation}
 C_{\mathrm t}(s)=
 \left\langle\widehat{\mathbf b}_{n}\cdot
 \widehat{\mathbf b}_{n+s}\right\rangle,
 \label{eq:tangent_correlation}
\end{equation}
with $\widehat{\mathbf b}_{n}=\mathbf b_n/|\mathbf b_n|$. Contour length,
end-to-end distance, tortuosity, and tangent correlations are standard
geometrical descriptors of semiflexible chains
\cite{HamprechtKleinert2005}; here they provide a quantitative test of the
polymer-like character of the finite clusters rather than assuming that
analogy from visual inspection alone \cite{Haddadi2021Langmuir}.

Cluster-size distributions were first accumulated independently for each
replica. In the particle-weighted distributions, an object of size $s$ was
weighted by $s$, so the resulting complementary cumulative distribution
gives the probability that a randomly selected particle belongs to an object
of size at least $s$. Fundamental-cycle distributions were number weighted,
whereas the displayed 2-core complementary distributions were weighted by
the number of particles in each core component. Per-replica distributions
were normalized before equal-weight averaging. Temperature-dependent cycle
and 2-core medians were evaluated with object weighting; a replica was
included only when at least 20 objects had been accumulated at that state
point. Particle weighting is particularly useful for cluster populations
because it reports the environment experienced by a randomly selected
particle, complementing object-weighted morphology statistics used for
competing-interaction colloids \cite{ValadezPerez2021}.

The long-time self-diffusion coefficient was determined from the
two-dimensional mean-square displacement,
\begin{equation}
\begin{split}
 \mathrm{MSD}(t)={}&
 \frac{1}{N}\sum_{i=1}^{N}
 \left\langle
 |\mathbf r_i(t_0+t)-\mathbf r_i(t_0)|^2
 \right\rangle_{t_0},\\
 D_{\mathrm{MSD}}={}&\lim_{t\rightarrow\infty}
 \frac{\mathrm{MSD}(t)}{4t}.
\end{split}
 \label{eq:diffusion_msd}
\end{equation}
Absolute particle coordinates were used and the MSD was averaged over all
available time origins. Lags up to one half of the production trajectory
were retained, and $D_{\mathrm{MSD}}$ was obtained from a linear fit to the
last half of this interval. Equation~\eqref{eq:diffusion_msd} is the
two-dimensional Einstein relation for self-diffusion
\cite{AllenTildesley2017}.

To connect local graph topology with mobility, one-frame displacements were
conditioned on the particle degree at the beginning of the interval:
\begin{equation}
 \delta_{k}^{2}=
 \left\langle|\mathbf r_i(t+\Delta t_{\mathrm{GSD}})
 -\mathbf r_i(t)|^2\right\rangle_{k_i(t)=k}.
\end{equation}
The topology-conditioned mobility contrasts reported in the main text are
\begin{equation}
 R_{\mathrm{end}}=\frac{\delta_1^2}{\delta_2^2},
 \qquad
 R_{\mathrm{branch}}=\frac{\delta_{\geq3}^2}{\delta_2^2}.
 \label{eq:mobility_ratios}
\end{equation}
For either topology-conditioned mobility ratio, a state point was
reported only when all four replicas contained, on average, at least two
particles of the corresponding conditioning class per stored configuration.
This population criterion avoids ratios constructed from effectively absent
endpoint or branch populations. Finally, bond
survival over the same interval was defined as
\begin{equation}
 C_{\mathrm{bond}}(\Delta t_{\mathrm{GSD}})=
 \frac{|E(t)\cap E(t+\Delta t_{\mathrm{GSD}})|}{|E(t)|}.
 \label{eq:bond_survival}
\end{equation}
Conditioning motion and bond persistence on network role follows the broader
strategy of relating mesoscale connectivity to heterogeneous dynamics in
colloidal networks \cite{ColomboDelGado2014}.

All observables were first averaged over the stored frames of each state
point. The four independent replicas were then combined with equal weight,
and statistical uncertainties are reported as the standard error of the
replica means.

\section*{S2. Robustness against the neighbor cutoff}
\label{sec:cutoff}

The graph representation requires a distance cutoff $r_{c}^{\rm graph}$ and the numerical
value of $\beta_1$ is not expected to be cutoff independent.  We therefore
repeated the graph construction for
\begin{equation}
 r_{c}^{\rm graph}=1.35,\;1.40,\;1.45,\;1.50,\;1.55,\;1.60,\;1.65 .
 \label{eq:S_cutoffs}
\end{equation}
The cutoff scan was evaluated every 20 saved trajectory frames.  Frame
averages were first calculated independently for each seed, after which the
four seeds were combined with equal weight.

\begin{figure}[t]
    \centering
    \includegraphics[width=\textwidth]
      {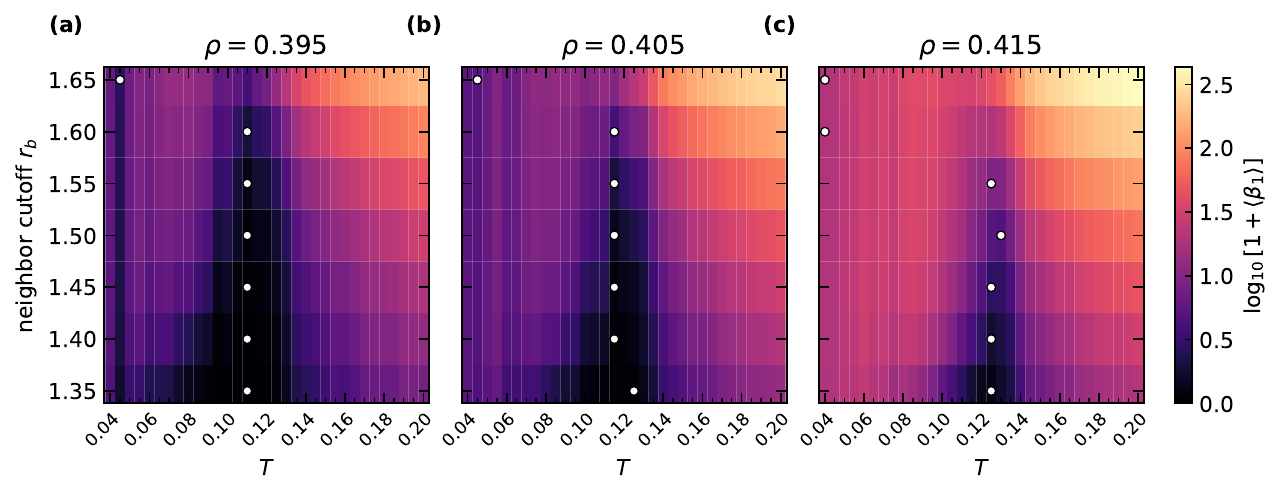}
    \caption{\label{fig:S_cutoff}
    Robustness of the first Betti number against the neighbor cutoff during
    heating for (a) $\rho=0.395$, (b) $\rho=0.405$, and
    (c) $\rho=0.415$.  The color scale represents
    $\log_{10}[1+\langle\beta_1\rangle]$.  White circles mark the temperature
    of the minimum at each cutoff.  Averages are formed with equal seed
    weight.  The persistence of the dark minimum over the first-neighbor
    cutoff range demonstrates that the nonmonotonic topological
    reconstruction is not an artifact of the particular choice
    $r_{c}^{\rm graph}=1.50$.}
\end{figure}

Figure~\ref{fig:S_cutoff} shows, for the three densities centered on the
optimal stripe-forming region,
$\log_{10}[1+\langle\beta_1\rangle]$ as a function of $T$ and $r_{c}^{\rm graph}$ during
heating.  The logarithmic color scale is used because increasing $r_{c}^{\rm graph}$
strongly increases the number of graph edges and hence the magnitude of
$\beta_1$.  White symbols identify the temperature of the minimum at each
cutoff.

For $\rho=0.395$, the minimum remains at $T=0.110$ throughout
$1.35\leq r_{c}^{\rm graph}\leq1.60$.  For $\rho=0.405$, it occurs at $T=0.115$ for
$1.40\leq r_{c}^{\rm graph}\leq1.60$, while the smallest cutoff shifts it modestly to
$T=0.125$.  For $\rho=0.415$, the minimum remains within
$T=0.125$--$0.130$ for $1.35\leq r_{c}^{\rm graph}\leq1.55$.  Thus, the
nonmonotonic reconstruction and its systematic density shift are robust
through the physically relevant first-neighbor range.

At the largest cutoffs, especially $r_{c}^{\rm graph}=1.65$, the minimum moves to the
low-temperature boundary for some densities.  This is not evidence that the
reconstruction disappears.  Rather, the graph becomes overconnected because
the cutoff begins to include contacts beyond the intended local
first-neighbor backbone.  The resulting additional edges generate many
short cycles already in the stripe phase and obscure the competition between
winding and nonwinding connectivity.  The main-text choice $r_{c}^{\rm graph}=1.50$ lies
well inside the robust interval.

The cutoff analysis supports two distinct conclusions.  The absolute value
of $\beta_1$ depends on the precise graph construction, as expected, whereas
the existence and density dependence of the topological minimum are stable
over a broad and physically motivated cutoff interval.

\section*{S3. Thermodynamic response along the heating isochores}
\label{sec:S_thermodynamics}

\begin{figure}[t]
    \centering
    \includegraphics[width=\textwidth]
      {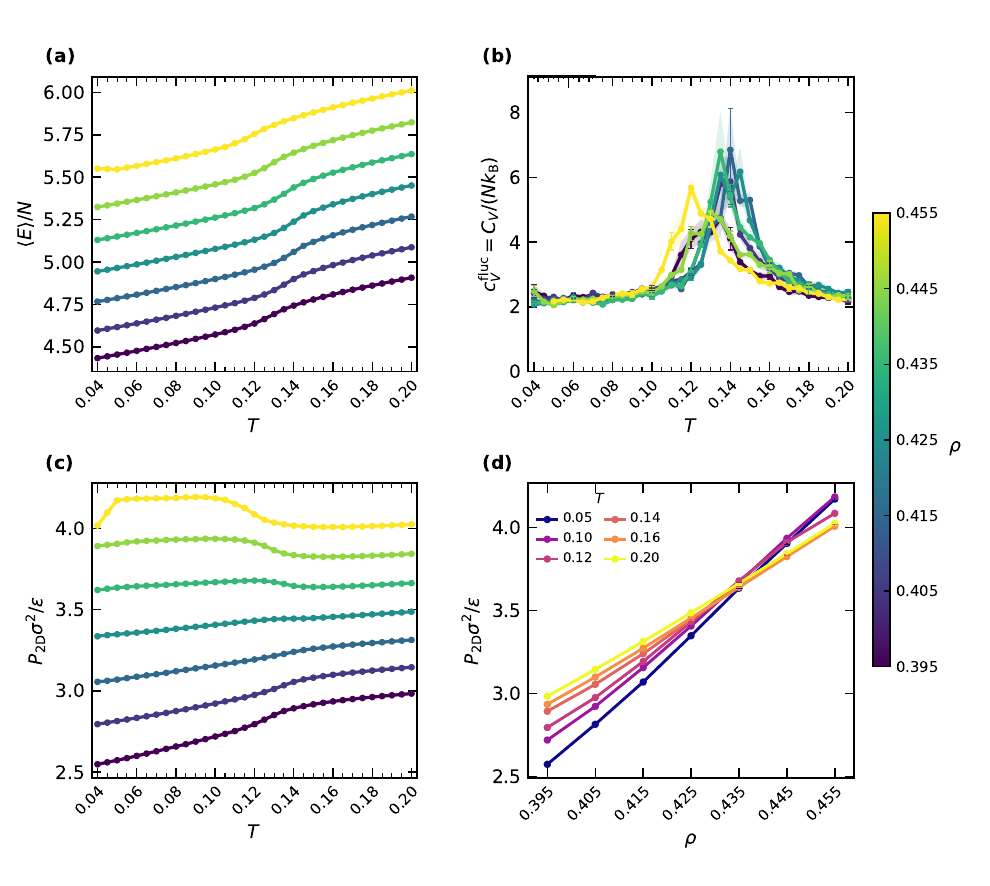}
    \caption{\label{fig:supp_thermodynamics}
  Thermodynamic evolution along the seven isochoric heating paths.
(a) Total energy per particle, $\langle E\rangle/N$.
(b) Isochoric heat capacity per particle obtained from the canonical
energy-fluctuation estimator,
$c_V^{\mathrm{fluc}}=[\langle E^2\rangle-\langle E\rangle^2]/(NT^2)$.
(c) Reduced in-plane pressure $P_{\mathrm{2D}}\sigma^2/\varepsilon$ as a
function of temperature. Colors in panels (a)--(c) identify density.
(d) Pressure--density representation for selected isotherms, showing the
reversal of their ordering at high density. Symbols show equal-replica means
at every retained state point; shaded bands denote standard errors over four
independent runs, with explicit error bars shown at every fourth temperature
in panels (a)--(c) for clarity.}
\end{figure}

We complement the structural and topological analysis with the total energy,
the isochoric heat capacity, and the in-plane pressure along the seven heating
isochores. At each state point, the instantaneous total energy and pressure
were sampled during the production trajectory. For each independent replica,
the mean energy per particle was calculated as
\begin{equation}
    e(T)=\frac{\langle E\rangle}{N}.
    \label{eq:S_energy_per_particle}
\end{equation}
The heat capacity per particle was calculated from the canonical
energy-fluctuation estimator
\begin{equation}
    c_V^{\mathrm{fluc}}(T)
    =\frac{\langle E^2\rangle-\langle E\rangle^2}{N T^2}
    =\frac{N\sigma_e^2}{T^2},
    \label{eq:S_cv_fluctuation}
\end{equation}
where $\sigma_e^2$ is the temporal variance of $E/N$ within one replica and
reduced units with $k_{\mathrm B}=1$ are used. As an independent consistency
check, we also evaluated the thermodynamic derivative
\begin{equation}
    c_V^{\mathrm{der}}(T)
    =\frac{\partial e(T)}{\partial T}.
    \label{eq:S_cv_derivative}
\end{equation}
At each temperature, a second-order polynomial was fitted to the five closest
points of the individual-replica energy curve, and its derivative at the
central temperature was retained. Near the limits of the temperature
interval, the procedure becomes a one-sided local fit. Both estimators were
evaluated separately for every replica before equal-weight ensemble
averaging. Their maxima agree to within one temperature increment,
$\Delta T=0.005$. Figure~\ref{fig:supp_thermodynamics} presents the
fluctuation estimator used in the main-text comparison.

The simulations were performed in a three-dimensional ESPResSo box of
thickness $L_z=10\sigma$, while the positions, velocities, and forces were
strictly confined to the $xy$ plane. The scalar pressure stored during the
simulations is one third of the trace of the volume-normalized pressure
tensor~\cite{Weik2019ESPResSo}. Since $P_{zz}=0$ in the present geometry, it
was converted to the two-dimensional areal pressure according to
\begin{equation}
 P_{\mathrm{2D}}
 =\frac{L_z}{2}\left(P_{xx}+P_{yy}\right)
 =\frac{3L_z}{2}P_{\mathrm{saved}}
 =15P_{\mathrm{saved}} .
 \label{eq:S_pressure_2d}
\end{equation}
The dimensionless pressure reported below is
$P_{\mathrm{2D}}\sigma^2/\varepsilon$. The conversion changes only the
pressure scale and leaves the temperature and density dependence unchanged.

The energy increases continuously along all seven isochores
[Fig.~\ref{fig:supp_thermodynamics}(a)], without a latent-heat-like jump at
the resolution of the present temperature grid. The heat capacity nevertheless
exhibits a broad maximum [Fig.~\ref{fig:supp_thermodynamics}(b)]. Its
characteristic temperature varies nonmonotonically with density: it increases
from the low-density side toward $\rho\simeq0.415$--$0.425$ and subsequently
decreases as the high-density, network-forming regime is approached. This
behavior is consistent with the density dependence of the orientational
response and susceptibility reported in the main text.

The heat-capacity maxima coincide closely with the interval in which the
global nematic order undergoes its principal decay. By contrast, the
topological reconstruction need not be centered on the same temperature.
At lower densities, the opening of winding stripe backbones and the minimum
of $\beta_1$ precede the thermodynamic maximum. At higher densities, winding
connectivity survives beyond orientational melting and becomes part of the
disordered network fluid. The thermodynamic anomaly is therefore associated
primarily with the collective loss of stripe alignment, whereas the topology
of the resulting fluid is selected over a broader density-dependent interval.

This separation reflects the different structural scales probed by energy
and graph topology. The energy is dominated by local pair separations and
coordination. In contrast, $\beta_1$ and periodic winding depend on how
locally similar filament segments connect across the simulation box.
Opening or reconnecting a comparatively small number of bonds can therefore
produce a substantial topological change while contributing only a weak
signal to the energy per particle.

The pressure uncovers an additional density-dependent thermodynamic response.
For $\rho=0.395$--$0.415$, $P_{\mathrm{2D}}$ increases monotonically with
temperature [Fig.~\ref{fig:supp_thermodynamics}(c)]. The increase becomes much
weaker at $\rho=0.425$, whereas the three highest densities exhibit a
nonmonotonic isochore. For $\rho=0.435$, $0.445$, and $0.455$, respectively,
the pressure has approximate maxima near $T=0.120$, $0.098$, and $0.091$,
followed by minima near $T=0.153$, $0.158$, and $0.160$. Consequently, the
thermal-pressure coefficient is negative over an extended interval that
contains the principal orientational and thermodynamic reconstruction.

The same reversal appears as a crossing of the isotherms in
Fig.~\ref{fig:supp_thermodynamics}(d). At every sampled temperature the
pressure remains monotonic in density, so that the finite-difference
isothermal stiffness is positive throughout the investigated interval.
Nevertheless, the ordering of the low- and high-temperature isotherms
reverses at high density. Comparing $T=0.05$ with $T=0.20$,
$P_{\mathrm{2D}}(0.20)-P_{\mathrm{2D}}(0.05)$ changes sign between
$\rho=0.435$ and $0.445$; linear interpolation places the crossing near
$\rho\simeq0.438$.

For a mechanically stable system,
\begin{equation}
 \left(\frac{\partial P}{\partial T}\right)_{\rho}
 =\frac{\alpha_P}{\kappa_T},
 \qquad
 \alpha_P=-\frac{1}{\rho}
 \left(\frac{\partial\rho}{\partial T}\right)_P ,
 \label{eq:S_thermal_pressure}
\end{equation}
where $\alpha_P$ and $\kappa_T$ are the isobaric thermal-expansion
coefficient and isothermal compressibility, respectively. The negative slope
of the high-density isochores therefore identifies a regime with
$\alpha_P<0$: at fixed pressure, the density increases rather than decreases
upon heating. Such water-like density anomalies are a characteristic
consequence of competition between local separations in core-softened
fluids~\cite{Oliveira06a, BordinBarbosa2018}, and their connection
with structural transformations and reentrant melting has previously been
demonstrated for the present two-length-scale model~\cite{Cardoso2021}.

Here, the anomalous interval overlaps the conversion of aligned stripes into
the high-density disordered network fluid. This is consistent with stripe
reconstruction relieving the packing frustration of the compressed
one-particle-thick morphology and redistributing local environments between
the two interaction scales. The pressure response thus provides a
thermodynamic counterpart to the density-dependent topology of the fluid.
The isochore crossing does not, however, determine a density discontinuity
between coexisting solid and liquid phases: the curves compare states at
different temperatures, and the present protocol does not locate a
solid--fluid coexistence line. We therefore interpret the result as a
density-anomalous thermal response accompanying the reconstruction, rather
than as direct evidence that the liquid is denser than the stripe solid at
coexistence.

\section*{S4. System-size dependence}
\label{sec:S_system_size}

\subsection*{Detailed comparison at the reference density}

To determine which features of the reconstruction survive changes in the
finite simulation box, we repeated the heating isochore at the optimal stripe
density, $\rho=0.405$, for $N=1024$ and $N=16384$.  These results are
compared with the $N=4096$ system used in the main text.  Four independent
realizations were analyzed for each size.  The box dimensions and the number
of initially prepared stripes were scaled with $N$ so that the density, the
longitudinal particle spacing, and the transverse stripe separation remained
unchanged.  All interaction, thermostat, equilibration, production, and
graph parameters were otherwise identical.

\begin{figure}[t]
    \centering
    \includegraphics[width=0.8\textwidth]
      {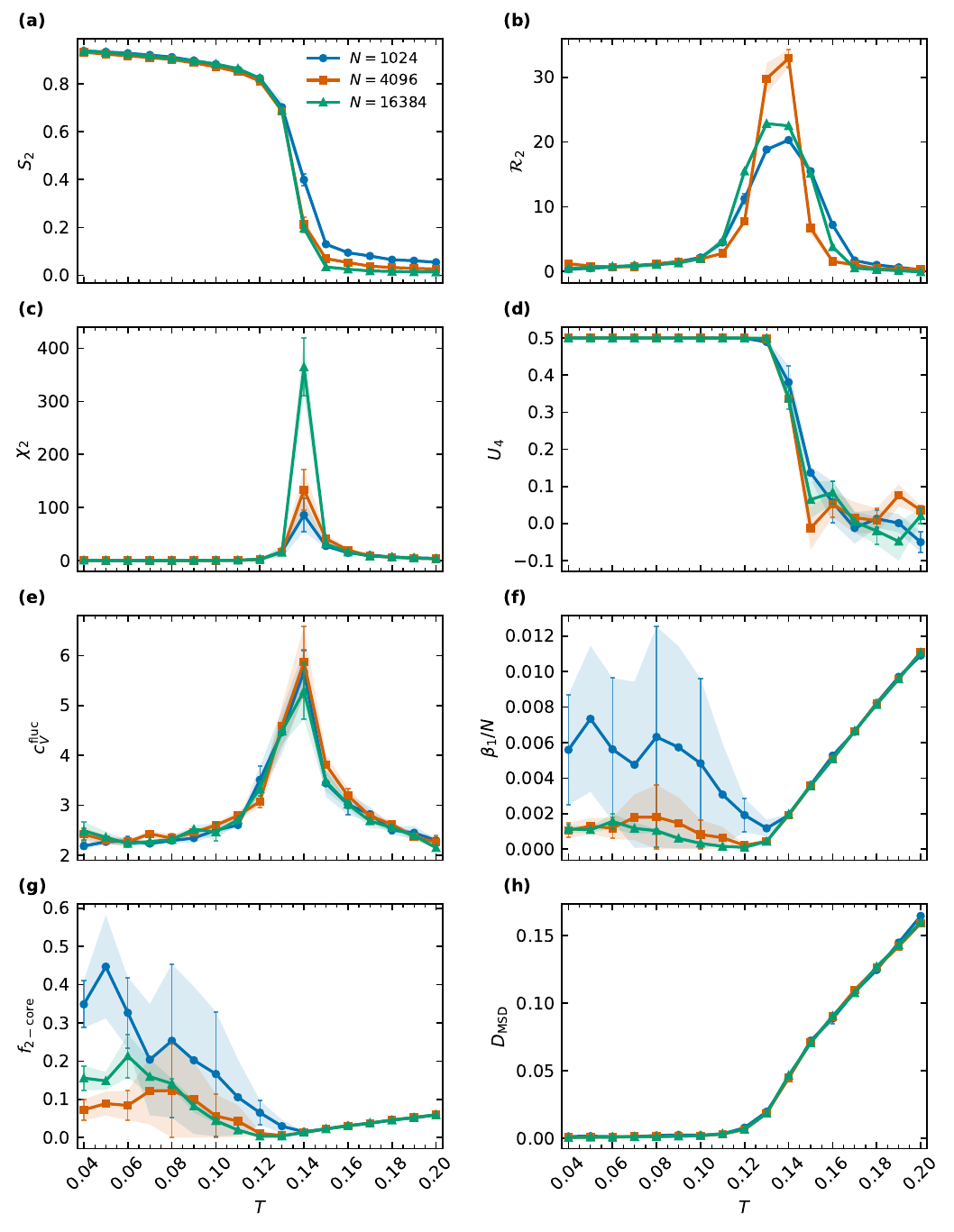}
    \caption{\label{fig:S_system_size}
    System-size dependence along the heating path at $\rho=0.405$.
    (a) Global nematic order parameter $S_2$.
    (b) Orientational response
    $\mathcal{R}_2=-\partial S_2/\partial T$.
    (c) Canonical orientational susceptibility
    $\chi_2=N[\langle S_2^2\rangle-\langle S_2\rangle^2]/T$.
    (d) Binder cumulant
    $U_4=1-\langle S_2^4\rangle/[2\langle S_2^2\rangle^2]$.
    (e) Heat capacity per particle obtained from energy fluctuations,
    $c_V^{\mathrm{fluc}}$.
    (f) Cycle density $\beta_1/N$.
    (g) Fraction $f_{\mathrm{2-core}}$ of particles belonging to the graph
    2-core.
    (h) Long-time diffusion coefficient $D_{\mathrm{MSD}}$.
    Symbols show equal-replica means at every temperature common to the
    three sizes. Shaded regions denote standard errors, with explicit error
    bars shown at every second temperature for clarity.}
\end{figure}

The additional simulations cover $0.04\leq T\leq0.20$ with
$\Delta T=0.01$, whereas the $N=4096$ data use
$\Delta T=0.005$.  Figure~\ref{fig:S_system_size} therefore displays only
temperatures shared by all three sizes.  We compare the global nematic order
$S_2$, its thermal response, the orientational susceptibility, the Binder
cumulant, the fluctuation heat capacity per particle, the cycle density
$\beta_1/N$, the 2-core fraction $f_{\mathrm{2-core}}$, and the long-time
diffusion coefficient $D_{\mathrm{MSD}}$.  These quantities either are
intensive or have been explicitly normalized by $N$.

To locate the temperature of the steepest orientational change without
assigning a discontinuity to a finite system, we define the orientational
response
\begin{equation}
    \mathcal{R}_2(T)=-\frac{\partial S_2}{\partial T}.
    \label{eq:S_orientational_response}
\end{equation}
For each replica, the derivative was evaluated by a quadratic local fit to
the five nearest temperatures; the resulting curves were subsequently
averaged with equal replica weight.  The quantity $\mathcal{R}_2$ is a
thermal slope, not a fluctuation susceptibility, and is used here only to
identify the interval in which orientational order changes most rapidly.

The principal decay of $S_2$ occurs over the same temperature interval for
all sizes [Fig.~\ref{fig:S_system_size}(a)].  The $N=4096$ and $N=16384$
curves are nearly superposed through the main decay, whereas the $N=1024$
system shows a somewhat broader crossover.  The residual high-temperature
value decreases with $N$, as expected for the finite-size magnitude of a
global order parameter in a disordered system.  This baseline effect does
not shift the principal disordering interval.

The corresponding orientational response is shown in
Fig.~\ref{fig:S_system_size}(b).  Its maximum occurs at $T=0.140$ for
$N=1024$ and at $T=0.135$ for $N=4096$.  For $N=16384$, the response has a
broad maximum extending from $T=0.13$ to $0.14$, with the two values nearly
indistinguishable on the scale set by $\Delta T=0.01$.  The three estimates
therefore locate the steepest loss of nematic order within the common
interval $T\simeq0.135$--$0.14$.

The frame-resolved fluctuations of the global nematic order provide a
complementary measure of this interval.  In reduced units, we define the
canonical orientational susceptibility as
\begin{equation}
    \chi_2(T)=\frac{N}{T}
    \left[
        \left\langle S_2^2\right\rangle
        -\left\langle S_2\right\rangle^2
    \right],
    \label{eq:S_orientational_susceptibility}
\end{equation}
where the moments are evaluated over the production frames of each replica.
The susceptibility is calculated separately for every replica before
performing the equal-weight replica average.  For all three sizes,
$\chi_2$ reaches its maximum at $T=0.14$
[Fig.~\ref{fig:S_system_size}(c)].  The peak values are
$86.0\pm31.2$, $133.4\pm38.1$, and $364.7\pm54.5$ for
$N=1024$, $4096$, and $16384$, respectively.  Thus, orientational
fluctuations grow markedly with system size, while the peak position remains
unchanged.  The increase is much weaker than the factor of $16$ separating
the smallest and largest particle numbers; with only three sizes and two
temperature resolutions, we do not assign a scaling exponent to this growth.

The distribution of $S_2$ is further characterized by the Binder cumulant
appropriate to the two-component nematic order parameter,
\begin{equation}
    U_4(T)=1-
    \frac{\left\langle S_2^4\right\rangle}
    {2\left\langle S_2^2\right\rangle^2}.
    \label{eq:S_binder_cumulant}
\end{equation}
As shown in Fig.~\ref{fig:S_system_size}(d), $U_4$ remains close to
$1/2$ in the ordered stripe regime and approaches zero after orientational
disordering, as expected when the two components of the global nematic order
fluctuate approximately as zero-centered Gaussian variables.  The decrease
becomes sharper with increasing $N$.  At $T=0.14$, the cumulants are
$0.381\pm0.045$, $0.336\pm0.008$, and $0.338\pm0.030$ for increasing
$N$.  The two larger systems are therefore statistically indistinguishable
at this temperature.  The curves do not, however, exhibit a single
well-resolved size-independent crossing, and the small negative values
observed after the loss of order are compatible with zero within their
uncertainties.  We consequently use $U_4$ as a distribution-shape and
finite-size diagnostic, rather than to assign a universality class or the
order of the transition.

The thermodynamic response independently identifies the same interval
[Fig.~\ref{fig:S_system_size}(e)].  For all three sizes,
$c_V^{\mathrm{fluc}}$ reaches its maximum at $T=0.14$, with peak values
$5.67\pm0.43$, $5.87\pm0.71$, and $5.29\pm0.56$ for increasing $N$.
The absence of systematic growth in the heat-capacity maximum contrasts with
the increasing orientational fluctuations.  Together with the absence of a
statistically significant negative Binder minimum and of latent-heat-like
behavior, this provides no evidence for a strongly first-order transition.
The stable peak positions of $\chi_2$ and $c_V^{\mathrm{fluc}}$, combined
with the strongest orientational slope, identify
\begin{equation}
    T_{\mathrm m}\simeq0.135\text{--}0.14
    \label{eq:S_melting_interval}
\end{equation}
as the principal orientational and thermodynamic melting interval.  No
additional energetic anomaly emerges when the linear dimensions are
increased.

The cycle density exhibits a distinct and earlier characteristic scale
[Fig.~\ref{fig:S_system_size}(f)].  For every size, $\beta_1/N$ decreases
as the winding stripe topology is removed, passes through a minimum, and
then increases when finite nonwinding loops proliferate.  The minima occur
at $T=0.130$, $0.115$, and $0.120$ for
$N=1024$, $4096$, and $16384$, respectively.  The two larger systems thus
place the topological minimum consistently near
\begin{equation}
    T_{\mathrm{topo}}\simeq0.12,
    \label{eq:S_topological_interval}
\end{equation}
below the orientational and heat-capacity maxima.  The smaller box broadens
the reconstruction and shifts the discrete minimum by one temperature
increment.  We therefore interpret $T_{\mathrm{topo}}$ as the characteristic
temperature of a topological crossover, rather than as a second
thermodynamic transition.

Exact low-temperature collapse of $\beta_1/N$ is not expected.  For an
ideal stripe array at fixed density, the number of independently winding
stripes grows with the linear box dimension, giving
\begin{equation}
    \frac{\beta_1^{\mathrm{wind}}}{N}
    \sim \frac{L}{N}
    \sim N^{-1/2}.
    \label{eq:S_winding_size_scaling}
\end{equation}
The winding contribution is therefore subextensive, whereas the local
nonwinding cycles in the polymer-like cluster fluid are extensive.  This
distinction accounts for the low-temperature offsets and the close collapse
of $\beta_1/N$ above the reconstruction interval.

The same scaling is visible directly in the low-temperature cycle length.
For the two larger systems, the median fundamental-cycle lengths at
$T=0.05$ are $68.75$ and $136.75$ particles for $N=4096$ and $16384$,
respectively.  If $n_{\parallel}$ is the initial number of particles along
a stripe, the normalized lengths are
\begin{equation}
    \widetilde{\ell}
    =\frac{\ell}{n_{\parallel}}
    =1.074\quad(N=4096),\qquad
    1.068\quad(N=16384).
    \label{eq:S_normalized_cycle_length}
\end{equation}
Thus, the low-temperature peak represents a contour that winds once around
the periodic box and scales with its linear dimension; it is not an
intrinsic microscopic cycle length.  In contrast, the high-temperature
cycle median is three particles for every size.  The particle-weighted
cluster distribution also converges in the fluid: at $T=0.20$ its median is
$14$ particles for all three sizes, while the upper deciles are $54.5$,
$56$, and $56$ particles for increasing $N$.

The 2-core fraction provides a complementary view of the reconstruction
[Fig.~\ref{fig:S_system_size}(g)].  Its quantitative size dependence is
largest in the stripe regime because a winding filament belongs entirely to
the 2-core, whereas a single opening can trigger recursive pruning along a
long chain.  Its minima occur at $T=0.14$, $0.13$, and $0.12$ for increasing
$N$.  Nevertheless, all sizes display the same decrease associated with the
loss of the winding backbone and the subsequent recovery produced by finite
cyclic structures.  The curves converge in the polymer-like fluid, showing
that the recovered 2-core is a local cyclic population rather than a
periodic-box artifact.

Finally, $D_{\mathrm{MSD}}$ is essentially size independent
[Fig.~\ref{fig:S_system_size}(h)].  The onset and rapid growth of diffusion
coincide for all sizes.  At $T=0.20$, for example, the diffusion
coefficients are $0.1644$, $0.1589$, and $0.1599$ for increasing $N$.
The near collapse of the two larger systems demonstrates that diffusive
fluidization is not controlled by the finite box.

Taken together, the system-size comparison at $\rho=0.405$ separates two
robust temperature scales. The topological minimum near
$T_{\mathrm{topo}}\simeq0.12$ marks
the conversion from winding backbones to predominantly open filaments,
whereas the maxima of the orientational susceptibility and thermodynamic
response near $T_{\mathrm m}\simeq0.14$ mark the principal loss of nematic
order. The available trends provide no evidence for a strongly first-order
transition, but the three sizes and the present temperature resolution are
insufficient to establish the transition order or a universality class.
Above this interval, the cycle density, 2-core population,
cluster-size distribution, and diffusion converge with increasing $N$.
Finite-size effects are concentrated in observables controlled by
system-spanning contours and modify their absolute scale, but not the
separation between the earlier topological crossover and the later
orientational melting at this density.

\subsection*{Density-resolved finite-size comparison}

We next test whether the density-selected reconstruction pathway survives
changes in the simulation box.  In addition to the detailed comparison at
$\rho=0.405$, simulations with $N=1024$, $4096$, and $16384$ are compared at
$\rho=0.425$, $0.435$, and $0.445$.  This representative set spans the
change, identified in the main text, from a fluid dominated by finite
polymer-like clusters to a fluid containing a persistent winding network.
Each point is an equal-weight average over four independent replicas.  The
density-resolved comparison is restricted to $T\leq0.18$, the largest
temperature interval with complete four-replica coverage for every displayed
pair $(N,\rho)$; it contains the full orientational and topological
reconstruction interval.

\begin{figure*}[t]
    \centering
    \includegraphics[width=\textwidth]
      {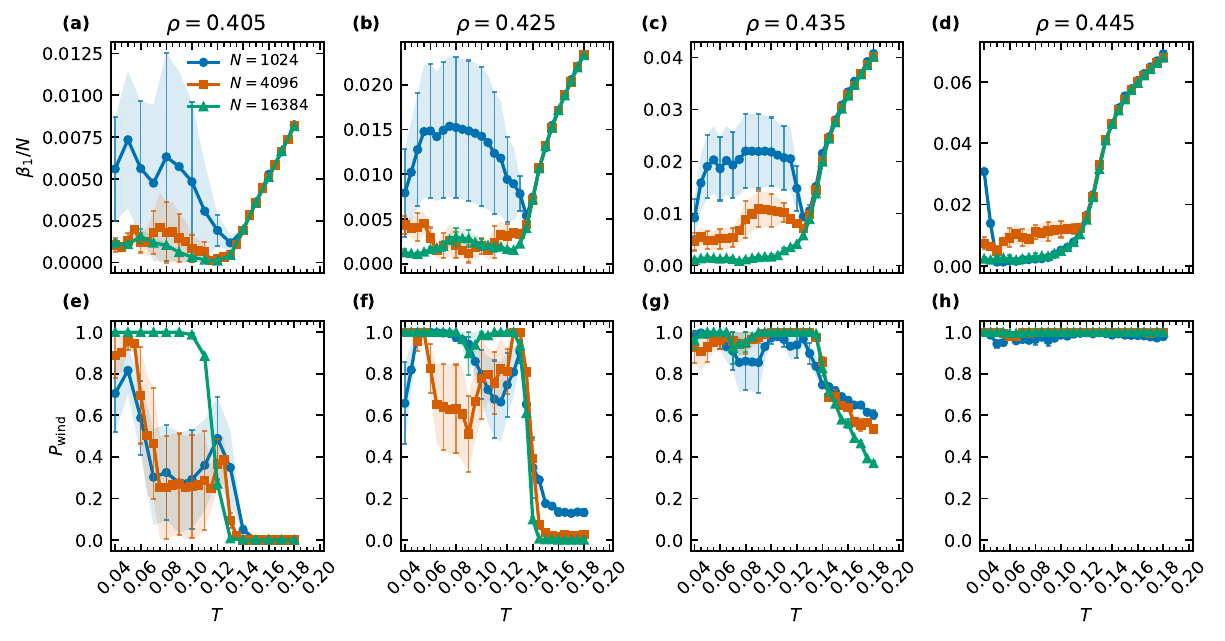}
    \caption{\label{fig:S_density_topology}
    Density-resolved system-size dependence of graph topology along the
    heating isochores. (a)--(d) Cycle density $\beta_1/N$ at
    $\rho=0.405$, $0.425$, $0.435$, and $0.445$, respectively.
    (e)--(h) Winding probability $P_{\rm wind}$ at the same densities.
    Symbols show equal-replica means over four independent runs. Shaded
    regions and explicit error bars denote standard errors; error bars are
    drawn at every second temperature for clarity. Only state points with
    complete four-replica production data are included.}
\end{figure*}

Figure~\ref{fig:S_density_topology} compares the cycle density and the
winding probability over this common range.  At $\rho=0.405$,
$\beta_1/N$ decreases as the winding stripe backbones open and subsequently
increases as finite local loops proliferate
[Fig.~\ref{fig:S_density_topology}(a)].  A corresponding minimum remains
visible at $\rho=0.425$, although its low-temperature amplitude and precise
position are more sensitive to $N$
[Fig.~\ref{fig:S_density_topology}(b)].  At $\rho=0.435$, the premelting
portion becomes still more size dependent, while the curves converge after
the rapid increase in the cycle population
[Fig.~\ref{fig:S_density_topology}(c)].  At $\rho=0.445$, no separated
interior minimum is resolved; instead, the growth associated with the
high-density winding network dominates the temperature dependence
[Fig.~\ref{fig:S_density_topology}(d)].  The nonmonotonicity of $\beta_1$
is therefore density dependent rather than a universal signature of stripe
melting.  Its low-temperature part contains subextensive winding cycles,
whereas the high-temperature collapse of $\beta_1/N$ demonstrates that the
newly formed local cycles constitute an extensive population.

The winding probability makes the change between the two fluid regimes
particularly transparent [Figs.~\ref{fig:S_density_topology}(e)--(h)].  At
$T=0.18$, $P_{\rm wind}$ is zero within resolution at $\rho=0.405$ for the
two larger systems and is only $1.2\times10^{-3}$ for $N=1024$.  At
$\rho=0.425$, it decreases from $0.132\pm0.018$ for $N=1024$ to
$0.026\pm0.006$ for $N=4096$ and zero for $N=16384$.  Thus, winding on the
low-density side is eliminated, rather than stabilized, as the linear box
dimension grows.  The intermediate density $\rho=0.435$ retains an
appreciable but size-dependent winding probability at $T=0.18$, with values
$0.602\pm0.021$, $0.534\pm0.015$, and $0.368\pm0.012$ for increasing $N$.
In contrast, at $\rho=0.445$ the winding probability remains at least
$0.979$ for all three sizes throughout the displayed high-temperature
range.  This persistence in the largest system rules out an interpretation
of the high-density network as a winding event generated only by a small
periodic box.  Because $P_{\rm wind}$ is a global binary observable, its
size dependence near the crossover is expected and we do not use it to
assign a sharp thermodynamic boundary between the two fluids.

\begin{figure}[t]
    \centering
    \includegraphics[width=\textwidth]
      {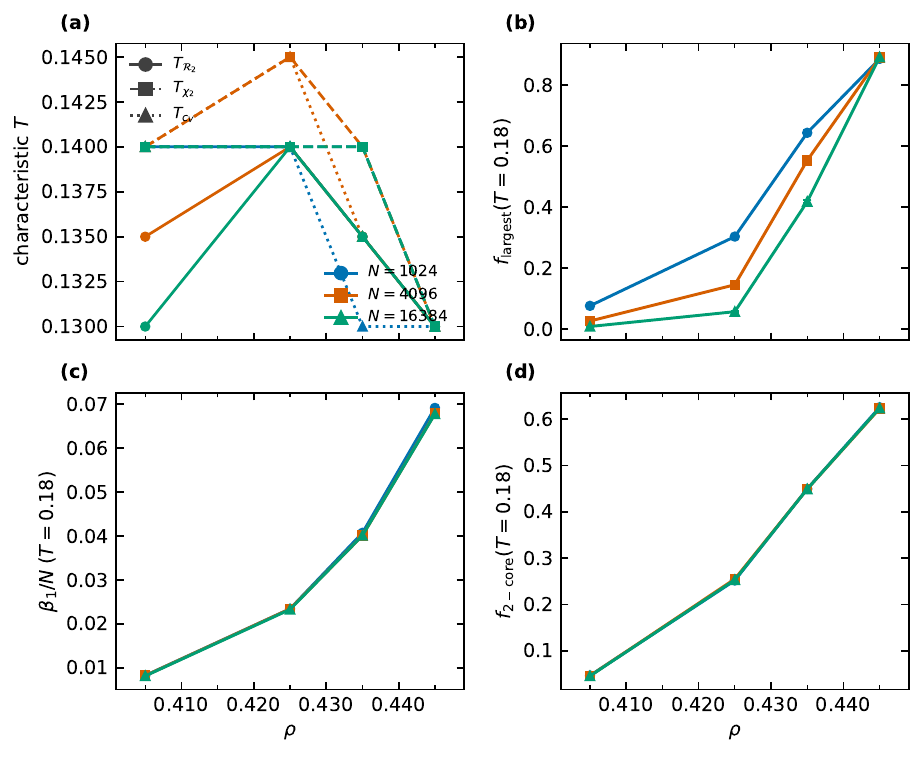}
    \caption{\label{fig:S_density_summary}
    Density-resolved summary of finite-size effects.
    (a) Temperatures of the maxima of the orientational response
    $\mathcal{R}_2$, orientational susceptibility $\chi_2$, and fluctuation
    heat capacity $c_V^{\rm fluc}$; the lines are guides to the eye and the
    temperature resolution is $0.005$--$0.01$.
    (b) Fraction of particles in the largest connected component at
    $T=0.18$. (c) Cycle density $\beta_1/N$ at $T=0.18$.
    (d) Fraction of particles in the graph 2-core at $T=0.18$.
    Colors and symbols identify $N$. Error bars in panels (b)--(d) are
    standard errors over four independent replicas and are smaller than the
    symbols when not visible.}
\end{figure}

The characteristic orientational and thermodynamic temperatures provide a
more conventional finite-size comparison.  Figure~\ref{fig:S_density_summary}(a)
shows the locations of the maxima of $\mathcal{R}_2$, $\chi_2$, and
$c_V^{\rm fluc}$.  For each density, the estimates from the three sizes agree
within one or two temperature increments.  The maxima lie near
$T=0.135$--$0.14$ at $\rho=0.405$, near $0.14$--$0.145$ at
$\rho=0.425$, between $0.13$ and $0.14$ at $\rho=0.435$, and near $0.13$
at $\rho=0.445$.  Their density dependence therefore survives increasing
$N$, whereas no systematic displacement with size is resolved.  We refrain
from assigning a single $T_{\rm topo}$ across this range because the
$\beta_1/N$ minimum ceases to be an interior, well-defined feature on the
high-density side.

The high-temperature structural measures reinforce this conclusion.  At
$T=0.18$, the largest-cluster fraction decreases strongly with increasing
$N$ at $\rho=0.405$, $0.425$, and $0.435$
[Fig.~\ref{fig:S_density_summary}(b)].  At $\rho=0.445$, by contrast, it is
essentially size independent: $f_{\rm largest}=0.886$, $0.890$, and $0.891$
for $N=1024$, $4096$, and $16384$, respectively.  The connected object in
this regime therefore contains an extensive fraction of the particles.
Simultaneously, both $\beta_1/N$ and $f_{\rm 2-core}$ collapse closely for
the three sizes at every displayed density
[Figs.~\ref{fig:S_density_summary}(c) and
\ref{fig:S_density_summary}(d)].  These intensive quantities show that the
local cyclic organization is already converged even where the probability
of global winding remains size dependent.

Taken together, Figs.~\ref{fig:S_density_topology} and
\ref{fig:S_density_summary} establish that the central conclusions are not
specific to $N=4096$.  The orientational and thermodynamic disordering
interval is stable, the high-temperature local topology is extensive, and
the high-density fluid retains a macroscopic connected backbone.  The main
finite-size corrections occur in observables tied explicitly to periodic
winding and to the largest connected object, as expected close to a
connectivity crossover.  The present evidence distinguishes a
finite-cluster fluid from a winding-network fluid but does not by itself
imply an additional thermodynamic phase transition between them.

\bibliography{apssamp}% Produces the bibliography via BibTeX.

\end{document}